\documentclass[a4paper,fleqn]{cas-sc}

\usepackage[numbers]{natbib}
\usepackage{xcolor}
\def\lazz{\mathrel{\mathchoice {\vcenter{\offinterlineskip\halign{\hfil
$\displaystyle##$\hfil\cr<\cr\sim\cr}}}
{\vcenter{\offinterlineskip\halign{\hfil$\textstyle##$\hfil\cr<\cr\sim\cr}}}
{\vcenter{\offinterlineskip\halign{
\hfil$\scriptstyle##$\hfil\cr<\cr\sim\cr}}}
{\vcenter{\offinterlineskip\halign{\hfil$\scriptscriptstyle##
$\hfil\cr<\cr\sim\cr}}}}}
\def\gazz{\mathrel{\mathchoice {\vcenter{\offinterlineskip\halign{\hfil
$\displaystyle##$\hfil\cr>\cr\sim\cr}}}
{\vcenter{\offinterlineskip\halign{\hfil$\textstyle##$\hfil\cr>\cr\sim\cr}}}
{\vcenter{\offinterlineskip\halign{
\hfil$\scriptstyle##$\hfil\cr>\cr\sim\cr}}}
{\vcenter{\offinterlineskip\halign{\hfil$\scriptscriptstyle##
$\hfil\cr>\cr\sim\cr}}}}}		
\def\pr{\prime}
\def\be{\begin{equation}}
\def\lan{\left\langle}
\def\ran{\right\rangle}
\def\ee{\end{equation}}
\def\barr{\begin{array}}
\def\earr{\end{array}}

\def\l{\left}
\def\r{\right}
\def\dis{\displaystyle}
\def\ed{\end{document}}
\def\f{\frac}

\def\cod{{\cal O}^\dagger}
\def\co{{\cal O}}

\def\cg{{\cal G}}

\def\cam{{\cal M}}

\def\ek{{\hat{E}_\kappa}}
\def\eh{{\hat{E}}}

\def\sh{\hat{\Sigma}}

\def\wm{{\widetilde {m}}}

\def\tmp{\widetilde{m_p}}
\def\tmn{\widetilde{m_n}}
\def\wm{{\widetilde {m}}}
\def\on{\overline{n}}

\def\ed{\end{document}}
\begin{document}

\shorttitle{}    

\shortauthors{V. K. B. Kota et al.}
\title [mode = title]{Embedded Random Matrix Ensembles to Statistical Shell Model: Operation of $q$-normal forms}  

\author[1]{V. K. B. Kota}%[orcid=0000-0000-0000-0000]
\cormark[1]
\ead{vkbkota@prl.res.in}
\cortext[1]{Corresponding author}
\affiliation[1]{organization={Physical Research  Laboratory},
	city={Ahmedabad},
	citysep={}, % Uncomment if no comma needed between city and postcode
	postcode={ 380 009}, 
	country={India}}

\author[2]{N. D. Chavda}[orcid=0000-0001-5958-1143]
\ead{ndchavda-apphy@msubaroda.ac.in}
\affiliation[2]{organization={Department of Applied Physics},
	addressline={Faculty of Technology and Engineering, The M.S. University of Baroda}, 
	city={Vadodara},
	citysep={}, % Uncomment if no comma needed between city and postcode
	postcode={ 390001}, 
	country={India}}

\author[3]{Manan Vyas}[orcid=0000-0003-2084-1109]
\ead{manan@icf.unam.mx }
\affiliation[3]{organization={Instituto de Ciencias F{\'i}sicas},
	addressline={Universidad Nacional Aut{\'o}noma de M{\'e}xico}, 
	city={Cuernavaca},
	citysep={}, % Uncomment if no comma needed between city and postcode
	postcode={62210}, 
	country={M{\'e}xico}}

%%%%%%%%%%%%%%%%%%%%%%%%%%	

\begin{abstract}
Embedded random matrix ensembles operating in nuclear shell model spaces, with nucleons occupying a finite set of single particle orbits and interacting via a two-body interaction, form the basis for statistical shell model. With sufficiently strong interaction, the level densities in shell model spaces take close to a Gaussian form and transition strength distributions close to a bivariate Gaussian form. In practice, partitioning via spherical configurations ($\widetilde{m}$) and angular momentum $J$ (also isospin where appropriate) are essential. The resulting statistical spectroscopy or statistical shell model was applied successfully in the past in some studies of nuclear level densities, orbit occupancies, $\beta$-decay matrix elements and so on. Going beyond these, recently it is recognized that embedded ensembles, in a better approximation, generate in-fact $q$-normal form ($q=1$ gives Gaussian and $q=0$ Wigner's semi-circle) for density of eigenvalues, bivariate $q$-normal form for transition strengths and conditional $q$-normal form for strength functions. These then allow us to develop statistical shell model with $q$-normal forms. These new developments in embedded ensembles and  statistical shell model are briefly reviewed in this paper. Also described, using some examples, is the role of the $q$ parameter in generating statistical properties of general quantum many-particle systems.
\end{abstract}

\maketitle

\section{Introduction}
\label{sec1}

We will begin with a condensed historical review of almost sixty years of Statistical Shell Model (SSM). However, those who are interested in first knowing the new perspective of the present article may skip the next three paragraphs of this introduction and go directly to the fourth paragraph.

The subject of statistical shell model (also called spectral distribution method or statistical nuclear spectroscopy) started with an article in 1967 by J.~B.~French where he showed that shell model energy ($E$) eigenvalue densities $\rho(E)$ (i.e. frequency function for $E$) are close to Gaussian in form \cite{jbf1}. Also, the partial densities defined over shell model spherical configurations ($\Gamma = \widetilde{m}$ or $\widetilde{m} J$ or $\widetilde{m} JT$ ) are Gaussians with small corrections. Note that, summing the partial densities $\rho^{\Gamma}(E)$ will generate level densities. These results are obtained using the then developed Rochester-Oak Ridge shell model code \cite{code1}. Construction of Gaussians is possible without using shell model matrices as the centroid $E_c$ and variance $\sigma^2$ defining a Gaussian will follow from the traces of the shell model Hamiltonian $H$ and that of $H^2$ respectively. Similarly, lower order corrections to the Gaussian follow from $H^3$ and $H^4$ traces. Note that a Gaussian $\rho_{\cg}(E) = (1/\sqrt{2\pi}\sigma)\exp -(E-E_c)^2/2\sigma^2$. Statistical spectroscopy or SSM with Gaussians (plus corrections) is possible due to the fact that traces of operators propagate from the defining spaces to the $m$-particle spaces with symmetries. Various developments and applications of SSM till about 2003 are described in detail in two books \cite{jbf2,jbf3} and in a conference proceedings \cite{jbf4}. Some major contributions in the development of SSM in the early years are as follows. (i) As Gaussians extend from $-\infty$ to $+\infty$, a method to determine the ground state energy $E_g$ is 
needed and Ratcliff produced a method \cite{jbf5} that is used by many groups in the applications of SSM. (ii) Trace propagation principles with configuration traces as an example are described in detail in \cite{jbf6}. (iii) Besides level densities, SSM theory for expectation values of an operator $K$ follow from the parametric differentiation of the eigenvalue density, i.e. the density generated by $H +\alpha K$ with $\alpha \rightarrow 0$. This gives a polynomial expansion (polynomials defined by $\rho(E)$) for expectation values \cite{jbf7}. More importantly,
this shows for example, orbit occupancies (expectation values of orbit
number operators) will be essentially linear in energy (polynomial expansion has simple convergence properties) and thus it is
possible to calculate with good accuracy ground state orbit occupancies. Examples for occupancies for neutrinoless double beta decay nuclei calculated using SSM and comparison with data are given recently \cite{jbf8}. (iv) SSM also gives a double polynomial expansion
for transition strength densities. Given a transition operator $\co$, the transition strength connecting an initial state with energy $E_i$ to a
final state $E_f$ is $\l|\lan E_f \mid \co \mid E_i\ran\r|^2$. Then, 
transition strength density is the transition strength multiplied by
the state densities at the two energies involved. Examination of the
convergence of the double polynomial expansion showed that the transition strength densities take bivariate Gaussian form and the correlation coefficient defining this bivariate Gaussian is given by the trace of
the operator $(\cod H \co H)$ \cite{jbf9}. Finally, the basis for the various distributions mentioned above lie in the operation of random matrix ensembles in shell model spaces and in particular the ensembles that include particle number, fermion character and interactions; see \cite{jbf10} and also \cite{jbf9,jbf11}. We will discuss this in more detail ahead.

Investigating from 1995 some aspects of spectral distributions using the OXBASH code \cite{code2,zel1}, in light of the developments in quantum chaos in nuclei and other finite quantum many-particle systems, Zelevinsky's group considered several improvements in the applications of SSM with particular reference to nuclear level densities. These studies include the following. (i)
Developed exponential convergence method for much better determination of the ground state and also advocated using shell model codes, where possible, for exact
determination of the ground state \cite{zel2}. (ii) Showed that direct construction of $\widetilde{m}J$ (where needed $JT$) partial densities
is much better than using $\widetilde{m}$ partial densities with Bethe's spin-cutoff factors for $J$ projection. Following this and using some of the earlier work on trace propagation, generated formulas for  $\widetilde{m}J$ centroids (trace of $H$) and variances (trace of $H^2$) that are good for machine calculations. More
importantly, recognized that the partial variances should be treated properly when multi-$\hbar \omega$ excitations are involved \cite{zel3}.
(iii) Developed a good method for exact removal of the center-of-mass spurious states from level densities \cite{zel4}. (iv) Recognized that it is important to remove the tails of the Gaussians by using finite range
Gaussians \cite{zel5}. Incorporating (i)-(iv), a high-performance algorithm to calculate spin- and parity-dependent nuclear level densities is developed and made available the associated computer codes \cite{zel5,zel6}. These codes are used in tabulating level densities of
$(2s1d)$ shell nuclei \cite{zel7} and also in some applications involving
$(2p1f)$ shell nuclei \cite{zel8}. All the work by Zelevinsky's group is
well summarized in \cite{zel9}; see also \cite{zel10}.

Although SSM approach is well developed for level densities, orbit occupancies and transition strengths (we will leave aside the improvements still needed in the theory for transition strengths), most important question is about the basis for SSM. Very early, it is recognized by French and Wong \cite{EE1} and Bohigas and Flores \cite{EE2} that random matrix theory (RMT) provides the basis but with the inclusion of the two-body nature of the nucleon-nucleon interaction. This gave rise to the introduction of two-body random matrix ensemble (TBRE) and shell model calculations with TBRE gave the ensemble averaged eigenvalue density to be Gaussian [classical Wigner-Dyson Gaussian orthogonal (GOE) and unitary (GUE) ensembles for example give semi-circle form \cite{EE3}) consistent with the shell model results generated by realistic effective interactions. Note that, in a TBRE the various two-particle $J$ matrices are represented by independent GOE's and then the two-body interaction defined by each member is used in shell model codes to generate the $H$ matrix for a given nucleus (for valence nucleons systems). With more general $k$-body interactions and for $m$ spinless fermions occupying $N$ number of single particle (sp) states, we have the so called fermionic embedded random matrix ensembles of $k$-body interactions FEE($k$). With a
GOE in $k$ particle spaces, we have in $m$ particle spaces FEGOE($k$). Similarly, a GUE in $k$ particle spaces generate FEGUE($k$). Mathematical definition of FEGOE($k$)/FEGUE($k$) is given in Section \ref{sec2} and their extension to bosonic ensembles BEGOE($k$)/BEGUE($k$) is straightforward. The FEGOE($k$) were investigated analytically for the first time by Mon and French \cite{jbf10} and they showed, using the so called binary correlation approximation, that as $k$ changes from 1 to $m$ the eigenvalue density changes from Gaussian form to semi-circle form. We will return to this important result in RMT in Section \ref{sec3}. Though the FEGOE($k$) were investigated further by Rochester group \cite{EE4,jbf9}, a major further development is in investigating FEGOE with two-body interactions in the presence of a mean-field, i.e. FEGOE(1+2). Analysis of FEGOE(1+2) ensembles (then the two-particle matrix generated by a two-body interaction is represented by a GOE and the mean-field one-body part is taken to be fixed in one particle space or it is also represented by an independent GOE in one particle spaces) showed that in the many-particle spaces,
as the strength of the two-body interaction increases, there is onset of
chaos and finally thermalization. To this end, the change in level statistics (from Poisson to GOE Wigner form), strength functions (changing from Breit-Wigner to Gaussian form), chaos measures such as information entropy and occupancy entropy, fidelity decay etc. are studied using both shell model examples and FEGOE(1+2) systems [also BEGOE(1+2) systems]. 
For details of these studies see \cite{jbf11,zel9,EE5,EE6,EE7,EE8} and references therein. In the shell model studies used are Rochester-Oak Ridge \cite{code1}, OXBASH \cite{code2}, NATHAN \cite{code3,EE5} and BIGSTICK \cite{code4} codes. The
FEGOE(1+2) results clearly established that nuclear interactions are strong enough to validate SSM principles \cite{zel10}.

Though it is well established that embedded random matrix ensembles (EE) provide the basis for the Gaussian forms used in SSM, more recently using a closer analysis of some of the analytical results for FEGOE($k$)/FEGUE($k$), it is recognized that EE indeed generate $q$-normal forms \cite{qn-1,qn-2,qn-3}. Therefore, the $q$-normal forms, instead of Gaussians, need to be
used in SSM \cite{qn-4}. Mathematical details of $q$-normal forms are given for example in \cite{qn-5,qn-6,qn-7,qn-8} and they are briefly discussed in Section \ref{sec2}. This new advancement in EE and SSM owes to the RMT results for the so called SYK model as derived by Verbaarschot and collaborators 
\cite{qn-9,qn-10,qn-10a,qn-11}. The SYK-RMT model involves Majorana fermions and it is a two parameter embedded random matrix ensemble. These are number of sp states $N$ and the body rank $k$ of the interactions. The classical ensemble GOE for example is a one parameter ensemble with the parameter being the matrix dimension $d$. In comparison, the basic EE($k$) are three parameter ensembles (see Section \ref{sec21}) with the parameters being number of sp states $N$, number of particles (fermions or bosons) $m$ and the body rank of the interactions $k$. Verbaarschot et al. \cite{qn-9,qn-10,qn-10a} proved analytically that the eigenvalue density for SYK-RMT is $q$-normal. This has opened up a new line of investigation of EE($k$)'s and their applications to SSM. To summarize, EE's do not generate Gaussian forms for SSM rather they generate the more general $q$-normal distributions. Correspondingly, eigenvalue densities, transition strength densities, and strength functions in SSM should be described by $q$-normal, bivariate $q$-normal, and conditional $q$-normal forms rather than ordinary Gaussians. With this recent recognition, the purpose of the present article is to present a short review of the results obtained so far for the 
$q$-normal forms in SSM. This review is needed as now there exists a number of new results ($q$-normal forms for fermion/boson ensembles, two-species extensions, strength-function and level-fluctuation analyses, extreme-value statistics, and a first attempt at a full $q$-normal-based SSM and so on) that are not available at one place, because the field is now poised to extend SSM beyond the old (1+2)-body Gaussian framework to realistic 3-body (and possibly 4-body) interactions and also as there are newer formal derivations showing that this theory is still actively developing. Now we will give a preview.

In Section \ref{sec2}, for completeness, we will introduce EGOE($k$)/EGUE($k$) for fermion and boson systems. Also presented are some of the important properties of $q$-normal, bivariate $q$-normal and conditional $q$-normal. In Section \ref{sec3}, binary correlation results for the moments of eigenvalue densities, transition strength densities and strength functions are presented and they show that these distributions follow $q$-normal, bivariate $q$-normal and conditional $q$-normal form respectively. Following these results, in Section \ref{sec4} results from further investigations of EE($k$) and EE(1+$k$) are presented. Section \ref{sec5} deals with $q$-normal and $q$-Hermite extensions of SSM. Finally, Section \ref{sec6} gives conclusions. 

\section{Preliminaries of embedded ensembles, $q$-normal distributions and $q$-Hermite polynomials}
\label{sec2}

\subsection{Embedded ensembles of $k$-body interactions with GOE/GUE embedding}
\label{sec21}

Given a system of $m$ particles (fermions or bosons) distributed in $N$ degenerate sp states and interacting via $k$-body $(1 \leq k \leq m)$ interactions, a class of embedded ensembles are generated by representing the $k$ particle Hamiltonian by a classical GOE/GUE and then the many-particle Hamiltonian ($m>k$) is generated by propagating the $k$ particle matrix elements to $m$ particle spaces using the Hilbert space geometry. In other words, $k$-particle Hamiltonian is embedded in the $m$-particle Hamiltonian and the non-zero $m$-particle Hamiltonian matrix elements are appropriate linear combinations of the $k$-particle matrix elements. Therefore, these random matrix ensembles are generically called {\it embedded random matrix ensembles or simply embedded ensembles}. Due to the $k$-body selection rules, many matrix elements of the $m$-particle Hamiltonian will be zero unlike in a GOE/GUE.

The random $k$-body Hamiltonian operator in second quantized form for a FEGOE/BEGOE ($\beta=1$) and FEGUE/BEGUE ($\beta=2$) is,
\be
H(k,\beta) = \dis\sum_{\alpha,\;\gamma} \; v^{\alpha,\gamma}_{k,\beta} \; \psi^\dagger(k; \alpha) \; \psi(k;\gamma) \;.
\label{eq1}
\ee 
Here, $\alpha$ and $\gamma$ are indices denoting $k$-particle states (configurations) in the occupation number basis. Distributing $k$ particles (fermions in agreement with Pauli's exclusion principle or bosons) in $N$ sp states will generate the complete set of these distinct configurations. The total number of these configurations are $\binom{N}{k}$ for fermions and $\binom{N+k-1}{k}$ for bosons. Operators $\psi^\dagger(k; \alpha)$ and $\psi(k;\gamma)$ respectively are $k$-particle creation and annihilation operators for fermions or bosons, i.e. $\psi^\dagger(k; \alpha) = \prod_{i=1}^{k} a^\dagger_{\mu_i}$, $\psi(k;\gamma) = \prod_{i=1}^{k} a_{\mu_i}$ for fermions and $\psi^\dagger(k; \alpha) =  {\cal N}_{\alpha} \; \prod_{i=1}^{k} b^\dagger_{\mu_i}$, $\psi(k;\gamma) = {\cal N}_{\gamma} \; \prod_{i=1}^{k} b_{\mu_i}$ for bosons. Here, ${\cal N}_{\alpha}$ and ${\cal N}_{\gamma}$ are the factors that guarantees unit normalization of $k$-particle bosonic states. The creation and annihilation operators $a^\dagger_p$ and $a_q$ for fermions ($b^\dagger_p$ and $b_q$ for bosons) satisfy the usual anti-commutation (commutation) relations for fermions (bosons). 

In Eq. (\ref{eq1}), $v^{\alpha,\;\gamma}_{k,\beta}$ matrix is chosen to be a $\binom{N}{k}$ [$\binom{N+k-1}{k}$] dimensional GOE (for $\beta=1$) or
GUE (for $\beta=2$) in $k$-particle spaces for fermion [boson] systems. That means $v^{\alpha,\;\gamma}_{k,\beta}$ are anti-symmetrized (symmetrized) $k$-particle matrix elements for fermions (bosons) chosen to be independent random Gaussian variables with zero mean and variance
\be
{\overline{v^{\alpha,\gamma}_{k,\beta} \; v^{\alpha^\prime,\gamma^\prime}_{k,\beta}}} = v^2 \; \left( {\delta_{\alpha,\gamma^\prime}} {\delta_{\alpha^\prime,\gamma}} + \delta_{\beta,1} {\delta_{\alpha,\alpha^\prime}} {\delta_{\gamma^\prime,\gamma}} \right) \;.
\label{eq2}
\ee 
Here and elsewhere in this article, the bar denotes 'ensemble averaging' and we choose $v=1$ without loss of generality. Now, distributing the $m$ fermions (bosons) in all possible ways in $N$ sp states generates all many-particle basis states defining $d_F(N,m)=\binom{N}{m}$ [$d_B(N,m)=\binom{N+m-1}{m}$] dimensional Hilbert space. Action of the Hamiltonian operator $H(k,\beta)$ defined by Eq. (\ref{eq1}) on the many-particle states generates the FEGOE($k$)/FEGUE$(k)$/BEGOE($k$)/BEGUE$(k)$ ensembles in $m$-particle spaces. Thus, these
EE will have three parameters $(N,m,k)$. However, if we include other degrees of freedom such as spin, then there will be more parameters; see \cite{EE7,EE8,zel10} for EE with additional degrees of freedom.

\subsection{$q$-normal distributions, moments and $q$ Hermite polynomials}
\label{sec22}

Let us begin with $q$ numbers $[n]_q$ defined by
\be
\l[n\r]_q = \dis\frac{1-q^n}{1-q} = 1+q + q^2 + \ldots+q^{n-1}\;.
\label{eq3}
\ee
Note that $[n]_{q \rightarrow1}=n$. Similarly $[n]_q! = \dis\Pi^{n}_{j=1} \,[j]_q$ with $[0]_q!=1$. Given these, the $q$-normal distribution $f_{qN}(x|q)$, with $x$ being a standardized variable (then $x$ is zero centered with variance unity), is defined as \cite{qn-5,qn-6,qn-7}
\be
f_{qN}(x|q) = \dis\frac{\dis\sqrt{1-q} \dis\prod_{k^\pr=0}^{\infty} \l(1-
q^{k^\pr +1}\r)}{2\pi\,\dis\sqrt{4-(1-q)x^2}}\; \dis\prod_{k^\pr=0}^{\infty}
\l[(1+q^{k^\pr})^2 - (1-q) q^{k^\pr} x^2\r]\;.
\label{eq4}
\ee
The $f_{qN}(x|q)$ function is defined in the range with
\be
S(q) = \l(x_{min} = -\dis\frac{2}{\dis\sqrt{1-q}}\;,\;x_{max}=+\dis\frac{2}{\dis\sqrt{1-q}}\r)
\label{eq5}
\ee
and $q$ takes values $0$ to $1$ in all the results presented in this review. Note that $f_{qN}(x|q) = 0$ outside $S(q)$ and the integral of $f_{qN}(x|q)$ over $S(q)$ is unity, $\int_{S(q)} f_{qN}(x|q)\,dx =1$.
For $q=1$, taking the limit properly
will give $f_{qN}(x|1)= (1/\sqrt{2\pi})\,\exp-x^2/2$, the Gaussian with $S(q=1)=(-\infty , \infty)$. Also, 
$f_{qN}(x|0)=(1/2\pi) \sqrt{4-x^2}$, the semi-circle with $S(q)=(-2,2)$. If we put back the centroid $\epsilon$ and the width $\sigma$ in $f_{qN}$, then $S(q)$ changes to
$$
S(q:\epsilon,\sigma) = \l(\epsilon -\dis\f{2\sigma}{\sqrt{1-q}}\;,\;\epsilon +\dis\f{2\sigma}{\sqrt{1-q}}\r)\;.
$$ 
The $q$-Hermite polynomials $He_n(x|q)$, that are orthogonal with $f_{qN}$ as the weight function, are defined by the recursion relation
\be
x\,He_n(x|q) = He_{n+1}(x|q) + \l[n\r]_q\,He_{n-1}(x|q)
\label{eq6}
\ee
with $He_0(x|q)=1$ and $He_{-1}(x|q)=0$. Note that for $q=1$, the $q$-Hermite
polynomials reduce to normal Hermite polynomials (related to Gaussian) and for
$q=0$ they will reduce to Chebyshev polynomials (related to semi-circle). 
The orthogonality property of $He_n(x|q)$'s is
\be
\dis\int^{2/\sqrt{1-q}}_{-2/\sqrt{1-q}} He_n(x|q)\,He_m(x|q)\,f_{qN}(x|q)\,dx = \l[n\r]_q!\,\delta_{mn}\;.
\label{eq7}
\ee
Using Eq. (\ref{eq7}), it is easy to derive formulas for the lower order reduced central moments $\mu_r(q)=\int_{Sq)}\,x^r f_{qN}(x)\,dx$ giving for example ($\mu_2(q)=1$ by definition),
\be
\barr{l}
\mu_4(q) = 2+q\;,\\
\mu_6(q) = 5+6q+3q^2+q^3\;,\\
\mu_8(q) = 14+28q+28q^2+20q^3+10q^4+4q^5+q^6\;.
\earr \label{eq8}
\ee
Note that all $\mu_{2r+1}(q)$ with $r=1,2,\ldots$ are all zero as $f_{qN}(x)$ is symmetrical in $x$.
See \cite{Kend} for various properties of moments, central moments, reduced central moments and cumulants [lowest two cumulants are the shape parameters skewness ($\gamma_1=\mu_3$) and excess ($\gamma_2=\mu_4-3$) parameters] of probability distributions.

Going further, the bivariate $q$-normal distribution $f_{biv-qN}(x,y|\xi , q)$, normalized to unity with
with $x$ and $y$ being standardized variables
and defined over $S(q)$ in both $x$ and $y$ spaces, is given by \cite{qn-6}
\be
\barr{l}
f_{biv-qN}(x,y|\xi , q) = f_{qN}(x|q) f_{qN}(y|q) h(x,y|\xi , q)\;;\\
\\
h(x,y|\xi ,q) \\
= \dis\prod_{k^\pr=0}^\infty \dis\frac{1-\xi^2 q^{k^\pr}}{
(1-\xi^2 q^{2k^\pr})^2 -(1-q)\,\xi\, q^{k^\pr}\,(1+\xi^2 q^{2k^\pr})\,xy +
(1-q)\xi^2 q^{2k^\pr} (x^2 +y^2)}\;,
\earr \label{eq9}
\ee
where $\xi$ is the bivariate correlation coefficient. Note that
$f_{qN}(x|q)$ and $f_{qN}(y|q)$ are the marginal densities of
$f_{biv-qN}$. Bivariate reduced central moments $\mu_{rs}(q) = \int_{S(q)} x^ry^s f_{biv-qN}(x,y|\xi , q)dxdy$ are symmetrical, i.e. $\mu_{rs}=\mu_{sr}$ and $\mu_{rs}=0$ for $r+s$ odd. Also $\mu_{20}=\mu_{02}=1$ and $\xi=\mu_{11}$. The $f_{biv-qN}$ can be expanded in terms of $q$-Hermite polynomials giving \cite{qn-6} (called Poisson-Mehler formula),
\be
f_{biv-qN}(x,y|\xi , q) = f_{qN}(x|q) f_{qN}(y|q) \l[\dis\sum_{n=0}^{\infty} \dis\f{\xi^n}{\l[ n\r]_q!}\,He_n(x|q) He_n(y|q)\r]\;.
\label{eq10}
\ee
Putting $\xi=1$ on both sides and operating over $S(q)$ gives,
\be
\delta(x-y) = f_{qN}(x|q) \l[\dis\sum_{n=0}^{\infty} \dis\f{1}{\l[n\r]_q!} He_n(x|q) He_n(y|q)\r] 
\label{eq11}
\ee
and this formula plays an important role in SSM. More importantly, Eq. (\ref{eq10}) allows us to derive formulas for $\mu_{rs}(q)$. For example,
the lower order bivariate moments $\mu_{rs}$ with $r+s=4$ and $6$ are,
\be
\barr{l}
\mu_{40}(q)=\mu_{04}(q)=2+q\;,\\
\mu_{31}(q)=\mu_{13}(q)= \xi\,(2+q)\;,\\
\mu_{22}(q) = 1 + \xi^2\,(1+q)\;,\\
\mu_{60}(q)=\mu_{06}(q)=5+6q+3q^2+q^3\;,\\
\mu_{51}(q)=\mu_{15}(q)=\xi\,\mu_{60}(q)\;,\\
\mu_{42}(q)=\mu_{24}(q)=(2+q) +\xi^2\,(3+5q+3q^2+q^3)\;,\\
\mu_{33}(q)=\xi\;(2+q)^2 + \xi^3\,(1+q)(1+q+q^2)\;.
\earr \label{eq12}
\ee

Given the bivariate $q$-normal $f_{biv-qN}$, the conditional $q$-normal densities ($f_{CqN}$) follow easily from Eq. (\ref{eq9}). Then, with
$h(x,y|\xi ,q)$ from Eq. (\ref{eq9}), we have
\be
\barr{l}
f_{CqN}(x|y; \xi , q)=f_{qN}(x|q) h(x,y|\xi ,q)\;,\\
f_{CqN}(y|x; \xi , q)=f_{qN}(y|q) h(x,y|\xi ,q)\;
\earr \label{eq13}
\ee
A very important property of $f_{CqN}$ that follows from Eq. (\ref{eq10}) is
\be
\dis\int_{S(q)} He_n(x|q) f_{CqN}(x|y; \xi ,q) dx = \xi^n He_n(y|q)\;.
\label{eq14}
\ee
From Eq. (\ref{eq14}), it is easy to infer that $f_{CqN}$ is normalized to unity over $S(q)$. It is important to recognize that the centroid $\epsilon(y:\xi ,q)$ of $f_{CqN}(x|y;\xi ,q)$ is not zero and its variance $\sigma^2(y:\xi ,q)$ is not unity. Using Eq. (\ref{eq14}) we have,
\be
\epsilon(y:\xi ,q) = \xi\,y\;,\;\;\;\sigma^2(y:\xi ,q) = (1-\xi^2)\;.
\label{eq15}
\ee
Thus, the centroid is linear in $y$ and variance independent of $y$. In addition, by writing $x^3$ and $x^4$ in terms of $q$-Hermite polynomials and using Eq. (\ref{eq14}) we have for the skewness and excess the formulas
\be
\barr{rcl}
\gamma_1(y:\xi ,q) & = & -(1-q)\dis\frac{\xi y}{\dis\sqrt{1-\xi^2}}\;,\\
\gamma_2(y:\xi ,q) & = & (q-1) +
(1-q)^2 \l[\dis\frac{\xi y}{\dis\sqrt{1-\xi^2}}\r]^2 + (1-q^2)\dis\frac{\xi^2}{(1-\xi^2)}\;.
\earr \label{eq16}
\ee
Thus, the $\gamma_1$ and $\gamma_2$ are zero only when $q=1$ (then the conditional distribution is a Gaussian and this result is well known \cite{Kend}). Let us mention that the so called Al-Salam-Chihara
polynomials are orthogonal with $f_{CqN}$ as the weight function \cite{qn-6}.

\section{Embedded ensembles and q-normal forms: Results from binary correlation approximation}
\label{sec3}

Turning to the eigenvalue densities (also transition strength densities and strength functions or local density of states)  generated by EE($k$), we will discuss in this Section results for FEGOE($k$)/FEGUE($k$) and point out their extensions to BEGOE($k$)/BEGUE($k$) where appropriate. It is important to recognize that the various densities considered in this Section will all be of same form for FEGOE($k$) and FEGUE($k$). Therefore, we will be using either of them though the final results apply
to both. 

\subsection{Eigenvalue density as $q$-normal}
\label{sec31}

The ensemble averaged eigenvalue density for FEGOE($k$)/FEGUE($k$) is deduced using the moment method supplemented by numerical verifications in \cite{qn-1}. Firstly, the moments are defined by $\overline {\lan H^p \ran^m}$ with $H$ given in Eq. (\ref{eq1}). By definition of the ensembles, the odd moments (p odd) will be zero and this includes the centroid. Formulas for the even moments to order 8 ($p=8$) are derived in \cite{jbf10} for FEGOE($k$) and in \cite{un1} for FEGUE($k$). Used here is
the dilute limit defined by $N \rightarrow \infty$, $m \rightarrow
\infty$, $m/N \rightarrow 0$ and $k/m$ fixed. Then, using the so called
binary correlation approximation \cite{jbf10,EE7}, the reduced moments up to order 8 for FEGUE($k$) are (with the variance or the second moment is given by $\overline {\lan H^2 \ran^m} = \binom{m}{k} \,\overline{\lan H^2(k) \ran^k}$) \cite{un1},
\be
\barr{l}
\mu_4(m,k) =  2 + f(m,k,1) \;,\\ 
\mu_6(m,k) =  5 + 6f(m,k,1) + 3\l[f(m,k,1)\r]^2 
+ f(m,k,2)f(m,k,1) \;,\\
\mu_8(m,k) = 14 + 28f(m,k,1) + 28\l[f(m,k,1)\r]^2 
 +  12\l[f(m,k,1)\r]^3 + 8f(m,k,2)f(m,k,1)\;, \\
 +  4f(m,k,1) \l[f(m,k,2)\r]^2 + 8 \l[f(m,k,1)\r]^2 f(m,k,2)\\
 +  f(m,k,1)f(m,k,2)f(m,k,3) +  2\l[f(m,k,1)\r]^2 \Delta\;; \\ 
f(m,k,r)={\dis\binom{m}{k}}^{-1}\;\dis\binom{m-rk}{k}\;.
\earr \label{eq17}
\ee
For the formula for $\Delta$ in Eq. (\ref{eq17}), see \cite{un1}.
Comparing these with the FEGOE($k$) formulas given in \cite{jbf10}, it is seen that
the moments to order 6 for FEGOE($k$) are same as those given in Eq.
(\ref{eq17}) and for $\mu_8$ only the last term is different.  Comparing the formulas in Eqs.(\ref{eq17}) and (\ref{eq8}), it is seen that the lower order reduced moments of FEGUE($k)$ and FEGOE($k$) will be essentially the same as those of the $q$-normal $f_{qN}(x)$ if we identify $q=f(m,k,1)$. The differences
in the 6th and 8th moments are verified to be a few percent or less. Significantly, as seen from the formula for $\mu_4(m,k)$, the $q$-parameter for FEGOE($k$)/FEGUE($k$) is given by
\be
q = \mu_4 -2 = {\dis\binom{m}{k}}^{-1}\;\dis\binom{m-k}{k} \sim 1-k^2/m \;.
\label{eq18}
\ee
Including finite $N$ corrections to $\mu_4$ as given in \cite{un2,un3,un4},
a better approximation for the $q$ parameter for FEGOE($k$)/FEGUE($k$) is,
\be
\barr{l}
q(N,m,k) = \dis\binom{N}{m}^{-1} \dis\sum_{\nu=0}^{min(k,m-k)}\; 
\dis\frac{\Lambda^\nu(N,m,m-k)\;\Lambda^\nu(N,m,k)\;d(g_\nu)}{
\l[\Lambda^0(N,m,k)\r]^2} \,; \\
\Lambda^\nu(N,m,r) =  \dis\binom{m-\nu}{r}\;\dis\binom{N-m+r-\nu}{r}\;,\;\;\;d(g_\nu)  = \dis\binom{N}{\nu}^2-\dis\binom{N}{\nu-1}^2\;.
\earr \label{eq19}
\ee
Note that $\overline{\lan H^2\ran^m} = \Lambda^0(N,m,k)$. Numerical calculations show that $q$ rapidly becomes $q=0$ as $k$ value increases. Note that $q=0$ for GOE/GUE and then we correctly have the semi-circle form for the eigenvalue density. Similarly, for $q=1$ we have Gaussian form as expected from the shell model with large $m$ values and using realistic interactions, TBRE and also EGOE(2) (then $k/m << 1$ and $q \sim 1$). For example, for $N=20$ and $m=8$,
Eq. (\ref{eq19}) gives $q=0.814$, $0.417$, $0.119$, 0.015 and $0$ for $k=1$, $2$, $3$, $4$ and $ \ge 6$, respectively. In addition, numerical embedded ensemble results for Gaussian to semi-circle transition are shown in
Fig. 1 and they are compared with the curve given by $f_{qN}(x)$ with $q$
defined by Eq. (\ref{eq19}). The agreement is excellent.

Turning to boson systems, here dense limit defined by $N \rightarrow \infty$, $m \rightarrow \infty$, $m/N \rightarrow \infty$ and $k/m$ fixed is important (dense limit will not exist for fermionic systems). Extension of the FEGOE($k$)/FEGUE($k$) results for the moments given by Eq. (\ref{eq17}) to BEGOE($k$)/BEGUE($k$) is available only for $\mu_4$ and it is proved using the so called $N \rightarrow -N$ law \cite{un4}. It is expected that the $q$-normal also applies to BEGOE($k$)/BEGUE($k$) in the dense limit and this is well
verified by numerical results in Fig. 2. It is important to recognize
that the finite $N$ formula for $\mu_4$ gives the formulas for $q$ for
BEGOE($k$)/BEGUE($k$) \cite{qn-1},
 \be
\barr{l}
q(N,m,k) = \dis\binom{N+m-1}{m}^{-1} \dis\sum_{\nu=0}^{\nu_{max}}\; \dis\frac{
\Lambda_B^\nu(N,m,m-k)\;\Lambda_B^\nu(N,m,k) \;d_B(g_\nu)}{\l[\Lambda_B^0(N,m,k)\r]^2}\;;\\
\Lambda_B^\nu(N,m,r) =  \dis\binom{m-\nu}{r}\;\dis\binom{N+m+\nu-1}{r}\;,\;\;\;
d_B(g_\nu)  = \dis\binom{N+\nu-1}{\nu}^2-\dis\binom{N+\nu-2}{\nu-1}^2\;.
\earr \label{eq20}
\ee
In Figs. 1 and 2, the values for $q$ given by finite $N$ formulas are used in constructing $f_{qN}(x)$. Eq. (\ref{eq20}) gives for example for a $(N=5, m=10)$ system the $q$ values to be 0.969, 0.861, 0.664, 0.405, 0.172,
0.045, 0.008, 0 for $k=1$, 2, 3, 4, 5, 6, 7, $\geq 8$.

Although we presented the results only for identical fermion (also identical boson) systems, for systems with two types of fermions (for example protons and neutrons) with H preserving the two fermion numbers,
it is seen that the $q$-normal form extend to the eigenvalue density for
two-fermion systems; to establish this, formulas for moments up to sixth order are derived in Ref. \cite{qn-14}. The $q$-normal is seen to extend also to
two species boson systems. The extension of the $q$-normal form to proton-neutron systems is clearly important for applications to SSM.

\begin{figure}
\centering
\includegraphics{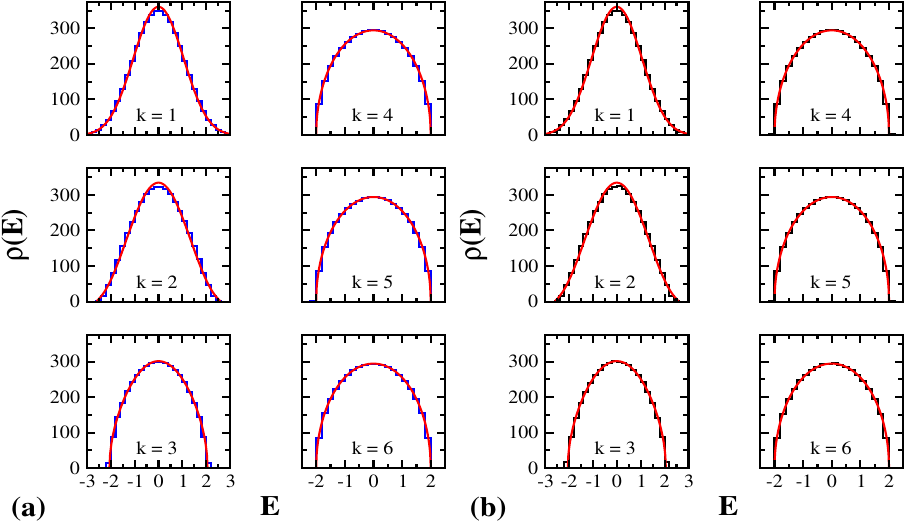}
\caption{Ensemble averaged state density $\rho(E)$ (histograms) for (a) FEGUE($k$) and (b) FEGOE($k$) as a function of standardized energy $\eh$ (denoted as $E$ in the figure). Used  in the calculations are 1000 members with $N=12$ and $m=6$ and $k$ changing from 1 to 6. The smooth curves [magenta in (a) and red in (b)] are $f_{qN}$ with $q$ defined by Eq.\eqref{eq19}. Figure is taken from \cite{qn-1}.}
\label{fig1-fk}
\end{figure}
\begin{figure}
\centering
\includegraphics{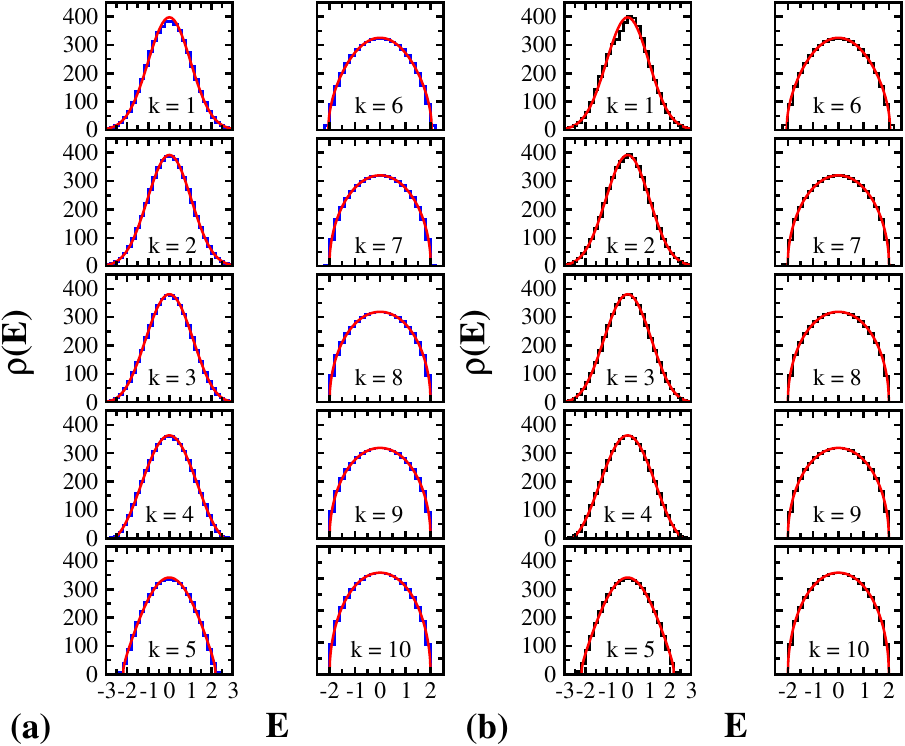}
\caption{Ensemble averaged state density $\rho(E)$ (histograms) for (a) BEGUE($k$) and (b) BEGOE($k$) as a function of standardized energy $\eh$ (denoted as $E$ in the figure). Used  in the calculations are 1000 members with $N=5$ and $m=10$ and $k$ changing from 1 to 10. The smooth curves [magenta in (a) and red in (b)] are $f_{qN}$ with $q$ defined by Eq.\eqref{eq20}. Figure is taken from \cite{qn-1}.}
\label{fig2-bk}
\end{figure}

\subsection{Transition strength density as bivariate $q$-normal}
\label{sec32}

Let us consider a system of $m$ spinless fermions in $N$ sp states with the Hamiltonian $H$ a $k$-body operator. Say $H$ generates the eigenstates
$\l|E\ran$ with $E$ denoting energy. Now, given a $t$-body transition operator $\co(t)$
acting on an eigenstate $\l|E_i\ran$ will populate the
state  $\l|E_f\ran$ with transition strength given by $\l|\lan E_f \mid \co \mid E_i\ran\r|^2$. The resulting bivariate transition strength density (normalized to unity) is,
\be
\rho_{biv-\co}(E_i,E_f) = \l[\lan \lan \cod \co \ran\ran^m\r]^{-1}\; 
\lan \lan \cod \delta(H-E_f) \co \delta(H-E_i)\ran \ran^m\;.
\label{eq21}
\ee
Note that $\lan \lan X \ran \ran^m =\sum_E \lan E \mid X \mid E\ran$. Our
interest is in deriving the statistical law for the transition strength densities as appropriate for nuclei. To this end,
$H$ is represented by FEGOE($k$)/FEGUE($k$) and the $\co$
by an independent FEGOE($t$)/FEGUE($t$). Then, formulas for the ensemble averaged (average with respect to both $H$ and $\co$ ensembles) bivariate
moments $\mu_{rs}(m,k,t)$ of $\overline{\rho_{biv-\co}(E_i,E_f)}$ are derived in \cite{qn-2} for $r+s \leq 6$ using the binary correlation approximation again. These formulas are
given in \cite{jbf9,jbf11,un5} and quite strikingly, they are close to those given by Eq. (\ref{eq12}) for $f_{biv-qN}$. We will describe this briefly below. Let us add that dividing the $\overline{\rho_{biv-\co}(E_i,E_f)}$ by the eigenvalue densities at $E_i$ and $E_f$ with give the ensemble averaged transition strengths. Also, the bivariate moments are 
defined by $\overline{\lan \cod H^s \co H^r\ran^m}$. Note that all odd moments  ($r+s$ odd) will be zero. Thus, the centroids are zero and the variances $\mu_{20} = \mu_{02} = \binom{m}{k}$. Similarly $\lan \cod \co\ran^m = \binom{m}{t}$. Now we will consider all other bivariate moments with $r+s=2$, 4 and 6.

Firstly, the bivariate correlation coefficient $\xi=\mu_{11}$ for ensemble averaged transition strength density is
\be
\mu_{11} = \xi= \l[\binom{m}{k}\r]^{-1}\;\dis\binom{m-t}{k} \;.
\label{eq22}
\ee
Similarly, the moments to order 4 and 6 and the $q$ parameter are (by using equations in \cite{jbf9,jbf11} and rewriting them in terms of $\xi$ and $q$ parameters),
\be 
\barr{rcl}
\mu_{40} = \mu_{04} & = & 2 + q\;;\;\;\; q = \l[\binom{m}{k}\r]^{-1}\,\binom{m-k}{k}\;,\\
\mu_{31} = \mu_{13} & = & \xi \mu_{40}\;,\\
\mu_{22} & = & 1 + \xi^2 (1+q)+ \xi (\Delta_0)\;,\\  
\mu_{60} = \mu_{06} & = & 5 + 6q + 3 q^2 + q^3 + q (\Delta_1)\;,\\
\mu_{51} = \mu_{15} & = & \xi\,\mu_{60}\;,\\
\mu_{42} = \mu_{24} & = & (2+q) + \xi^2 (3 + 5 q + 3 q^2 + q^3) + \xi (X)\;,\\
\mu_{33} & = & \xi \l[4 + 4q + q^2\r] + \xi^3 \l[ 1 + 2q 
+2q^2 + q^3\r] + \xi (Z)\;.
\earr \label{eq23}   
\ee
Formulas for $\Delta_0$, $\Delta_1$, $X$ and $Z$ are given in \cite{jbf9,jbf11}. Numerical calculations are used to verify that for some typical values of $(N,m,k,t)$, these are indeed $\sim 0$. With this, by comparing Eq. (\ref{eq23}) with Eq. (\ref{eq12}), it is clear that 
the transition strength densities generated by FEGOE/FEGUE are well represented by the bivariate $q$-normal distribution. Thus, by changing
$E_i$ and $E_f$ by the corresponding standardized variables $x$ and $y$
respectively,
$$
\overline{\rho_{biv-\co}(x,y)} \rightarrow f_{biv-qN}(x,y)
$$
with $\xi$ and $q$ given by Eqs. (\ref{eq19}) and (\ref{eq22}).
Let us mention that it is well known in statistics \cite{Kend} and in random matrix theory \cite{EE4,jbf3} that lower order moments generate the form of a probability distribution. 

Eq. (\ref{eq19}) gives the formula for the $q$ parameter with finite $N$ corrections. Similarly, formula with finite $N$ corrections to the correlation coefficient $\xi$ is \cite{qn-2}
\be
\xi(t,k) = \dis\sum_{\nu=0}^{min(t,m-k)}\; 
\dis\frac{\Lambda^\nu(N,m,m-t)\;\Lambda^\nu(N,m,k)\;d(g_\nu)}{\binom{N}{m}\;
\Lambda^0(N,m,k)\;\Lambda^0(N,m,t)} \,.
\label{eq24}
\ee
Note that the function $\Lambda$ is defined in Eq. (\ref{eq19}). For example, for $(N,m)=(20,10)$ the values for $\xi(1,k)$ are 0.682, 0.559,
0.455, 0.364, 0.152 for $k=2$, 3, 4, 5 and 8 respectively. Similarly,
for $\xi(2,k)$ they are 0.465, 0.314, 0.21, 0.136 and 0.027 respectively.  

Although we have restricted to transition operators that preserve $m$ (then $E_i$ and $E_f$ belong to the same system) in the discussion above, it is also possible to analyze $\mu_{rs}$ with $r+s=4$ and $(rs)=(11)$ for beta and neutrinoless double beta decay type operators and also for particle removal operators using the results in \cite{un5}. More importantly, they will also give formulas, with finite $N$ corrections, for $\xi$ and $q$ for the transition strength densities generated by these operators. Again, these are important for applications of bivariate $q$-normal for transition strength densities in SSM.
  
\subsection{Strength functions as conditional q-normal}
\label{sec33}

Wavefunction structure in finite quantum many-body systems such as atomic nuclei follows from the form of the strength functions. Given the eigenstates expanded in terms of a set of physically motivated basis states, strength functions correspond to the spread of a basis state over the eigenstates. They also correspond to partial densities in SSM; for
a more general discussion of strength functions see \cite{jbf11}. By
deriving analytical formulas, using EE, for the lowest four moments of the strength functions, it is shown in \cite{qn-3} that the conditional $q$-normal$f_{CqN}$ to a good approximation represents strength functions. For good numerical tests of this result, see \cite{qn-12,qn-13}. Here below we will describe the EE formulas and important structures they display.

Let us consider a system of $m$ fermions in $N$ single particle (sp) states and the Hamiltonian $H$ for the system is say, 
\be
H = H_0(t) + \lambda V(k)
\label{eq25}
\ee
where $H_0$ is a $t$-body operator, $V$ is a $k$-body operator and $\lambda$ is the strength parameter. We will assume that $t << k$ and for $m$ fermions, obviously $k \leq m$. In many physical applications $t=1$ with $H_0$ representing a mean-field one-body Hamiltonian \cite{EE7}. 
Now, strength functions form is studied by considering the structure of eigenfunctions of $H$ expanded in terms of the unperturbed $H_0$ eigenstates (basis states). Denoting $\l|\kappa, \alpha\ran$ as the eigenstates of $H_0$ forming a complete set with $H_0 \l|\kappa, \alpha\ran = E_\kappa \l|\kappa, \alpha\ran$ and $\l|E, \beta\ran$ as the eigenstates of $H$ forming a complete set with $H\l|E, \beta \ran = E \l|E, \beta \ran$ (with $\alpha$ and $\beta$ labeling the respective degeneracies in $H_0$ and $H$ spectrum), we can expand the eigenstates of $H_0$ in the eigenbasis of $H$ giving,
\be
\l|\kappa, \alpha\ran = \sum_{E, \beta} C_{\kappa, \alpha}^{E, \beta} \l|E, \beta \ran \;.
\label{eq26}
\ee
Now, the strength function $F_\kappa(E)$ is,
\be
\barr{rcl}
F_\kappa(E)  & = & \lan \delta(H-E)\ran^{\kappa}\;;\\
& = & \dis\f{1}{d\rho_1(E_\kappa))} \dis\sum_{\alpha \in \kappa , \beta \in E}
\l|C_{\kappa , \alpha}^{E , \beta}\r|^2 \;.
\earr \label{eq27}
\ee
Here, $d$ is $m$ particle space dimension, $d\rho_1(E_\kappa)$ gives number of $H_0$ states with same basis state energy $E_\kappa$ and similarly $d\rho_2(E)$ gives number of eigenstates of $H$ with same eigen energy $E$. Note that $\rho_2(E) = \lan \delta(H-E)\ran^m$ is the eigenvalue density generated by $H$ and similarly, $\rho_1(E_\kappa) = \lan \delta(H_0 - E_\kappa) \ran^m$ is the eigenvalue density generated by $H_0$. 

In order to derive the form of $F_\kappa(E)$, formulas for the ensemble averaged lower order moments of $F_\kappa(E)$ are derived by representing $H_0(t)$ by FEGOE($t$)/FEGUE($t$) and $V(k)$ by FEGOE($k$)/FEGUE($k$) and assuming that they are independent. From now on, for brevity, often we will drop $t$ in $H_0$ and $k$ in $V$.
By definition, we have $\overline{\lan H_0\ran^m}=0$ and $\overline{\lan H\ran^m}=0$. These are the centroids of $\overline{\rho_1(E_\kappa)}$
and $\overline{\rho_2(E)}$ respectively. Similarly, the corresponding variances are $\sigma_{H_0}^2 = \overline{\lan H_0^2\ran^m}$ and
$\sigma_{H}^2 = \overline{\lan H^2\ran^m} =\overline{\lan H_0^2\ran^m} +
\lambda^2 \overline{\lan V^2\ran^m}$. Scaling the eigenvalues $E$ with their width $\sigma_H$, the moments of $F_\kappa(E)$ are given by
\be
M_r(E_\kappa) = \l[\sigma_H^r\r]^{-1}\;\overline{\lan H^r \ran^\kappa}\;.
\label{eq28} 
\ee
It is important to note that $\lan H_0^p \ran^{\kappa} = E^p_\kappa$ as $\kappa$ are eigenstates of $H_0$ with eigenvalues $E_\kappa$. Therefore,
$\lan H^r\ran^\kappa = \lan H^r\ran^{E_\kappa}$ is an expectation value and it can be written in terms of the polynomials generated by $\overline{\rho_1(E_\kappa)}$. For a general operator $K$, the polynomial expansion to second order (usually this is sufficient) is\cite{jbf7,jbf3,qn-3},
\be
\barr{l}
\lan K \ran^{\kappa} = \dis\sum_\mu \lan K P_\mu(H_0) \ran^m P_\mu(E_\kappa)\;;\\
P_0(x) = 1,\;\;\;P_1(x) = \hat{x},\;\;\;P_2(x) = \dis\frac{(\hat{x})^2-1}{\dis\sqrt{\mu_4-1}}\;.
\earr \label{eq28a}
\ee
Note that $\hat{x}=E_{\kappa}/\sigma_{H_0}$ and $\mu_4$ is the fourth reduced moment of $\overline{\rho_1(E_\kappa)}$. 
Before proceeding further, let us mention that the moments $M_r$ in Eq. (\ref{eq28}) give the central moments $\cam_r$ to be (for $r \le 4$),
\be
\barr{rcl}
\cam_2 & = & M_2 - M_1^2\;,\\
\cam_3 & = & M_3 - 3M_2 M_1 +2M_1^3 \;,\\
\cam_4 & = & M_4 - 4M_3 M_1 +6M_2 M_1^2 -3M_1^4\;.
\earr \label{eq29}
\ee
Without going into details (see \cite{qn-3} for the details) we will now give the formulas for $M_r$, $r=1-4$. These will then give the variance
$\sigma^2(E_\kappa)$, skewness $\gamma_1(E_\kappa)$ and excess $\gamma_2(E_\kappa)$ via Eq. (\ref{eq29}). Firstly the centroid $M_1(E_\kappa)$ is
\be
M_1(E_\kappa) = \l[\sigma_H\r]^{-1} \;\overline{\lan H \ran^{\kappa}} = \l[\sigma_H\r]^{-1}\;\overline{\lan H_0\ran^{\kappa}} = \xi\, \ek\;.
\label{eq30}
\ee
Note that the ensemble average of $\lan (H_0)^r [V(k)]^s \ran^\kappa$ is zero for any $r$ and any odd $s$ and this is
used above (for $r=0,s=1$). More importantly, $\xi$ is a correlation coefficient and clearly (with $\ek=E_\kappa/\sigma_{H_0}$),
\be
\xi=\sigma_{H_0}/\sigma_H = \dis\sqrt{\dis\frac{\binom{m}{t}}{\binom{m}{t} + \lambda^2 \; {\binom{N}{t}}^{-1}\,\binom{N}{k}\;\binom{m}{k}}}\;.
\label{eq31}
\ee
Now, the second moment $M_2(E_\kappa)$ is (absorbing $\lambda$ in $V(k)$ and using $\sigma_H^2=\sigma_{H_0}^2+\sigma_V^2$),
\be
\barr{rcl}
M_2(E_\kappa) & = & \l[\sigma^2_H\r]^{-1} \overline{\lan H^2 \ran^\kappa} = \l[\sigma^2_H\r]^{-1}\;\overline{\lan H^2_0 + V^2 + (H_0V + VH_0) \ran^{\kappa}} = \xi^2 (\ek)^2 + (1-\xi^2) \\ 
\Rightarrow \cam_2(E_\kappa) & = & (1-\xi^2) \;.
\earr \label{eq31a}
\ee
Thus, the strength function centroid is linear in $E_\kappa$ (with slope $\xi$) and the variance $\cam_2(E_\kappa)$ is independent of $E_\kappa$. Now let us consider $M_3(E_\kappa)$,
\be
\barr{l}
M_3(E_\kappa) = \l[\sigma^3_H\r]^{-1} \;\overline{\lan H^3 \ran^{\kappa}} \\
= \l[\sigma^3_H\r]^{-1}\;\overline{\lan H_0^3 + V^3 + (H_0 V^2 + V^2H_0) + (H^2_0 V + VH^2_0) + VH_0V + H_0 V H_0 \ran^{\kappa}} \\
= \xi^3 (\ek)^3 + 2 \xi (1-\xi^2)\ek + \l[\sigma^3_H\r]^{-1}\;
\overline{\lan V H_0 V\ran^{\kappa}}\;.
\earr \label{eq32}
\ee
Now evaluating the last term using only the first two terms in Eq. (\ref{eq28a}), and simplifying Eqs. (\ref{eq32}) and (\ref{eq29}) will give the remarkable formula \cite{qn-3},
\be
\mu_3(\ek) = \gamma_1(\ek) = - \dis\f{\xi \l(1-q^{hv}\r) \ek}{\dis\sqrt{1-\xi^2}} \;.
\label{eq33}
\ee
Here, the $q^{hv}$ is
\be
q^{hv} = \dis\f{\overline{\lan H_0 V H_0 V\ran^m}}{\sigma_{H_0}^2 \sigma_V^2} = {\binom{m}{k}}^{-1}\;\binom{m-t}{k}\;.
\label{eq34}
\ee
Therefore, the first three moments $M_1(E_\kappa)$, $\cam_2(E_\kappa)$ and $\mu_3(E_\kappa)$ of the strength function from EE are same as those of
$f_{CqN}(y)$ with $y=\ek$ and $\xi$ and $q$ given by Eqs. (\ref{eq31}) and (\ref{eq34}) respectively; see Eqs. (\ref{eq15}) and (\ref{eq16}).
Finally, the fourth moment is given by 
\be
M_4(E_\kappa) = \l[\sigma^4_H\r]^{-1} \;\overline{\lan H^4 \ran^{\kappa}} =
\l[\sigma^4_H\r]^{-1} \;\overline{\lan\l(H^2_0 + V^2 + (H_0V + VH_0)\r)^2 \ran^{\kappa}}\;.
\label{eq35}
\ee
Using Eq. (\ref{eq28a}) and carrying out simplifications as described in
detail in \cite{qn-3}, we have
\be
\mu_4(\ek) \simeq \mu_4^0(\ek) =  \l(2+q^{hv}\r) + \dis\f{\xi^2 (\ek)^2 \l(1-q^{hv}\r)^2 + \xi^2 \l[1-\l(q^{hv}\r)^2\r]}{1-\xi^2}\;.
\label{eq36}
\ee
with $\gamma_2(\ek) = \mu_4(\ek)-3$, it is seen by comparing Eq. (\ref{eq36}) with Eq. (\ref{eq16}) that the strength functions from EE
are well approximated by $f_{CqN}(E|y;\xi, q^{hv})$ with $y=\ek$ and $\xi$ and $q^{hv}$
given by Eqs. (\ref{eq31}) and (\ref{eq34}) respectively. It is useful
to note that the finite $N$ formula for $q^{hv}$ is \cite{qn-3},
\be
q^{hv} = \dis\frac{\dis\sum_{\nu=0}^{min(t,m-k)}\,
\Lambda^{\nu}(N,m,k)\, \Lambda^{\nu}(N,m,m-t)\,d(g_\nu)}{
\dis\binom{N}{m}\;\Lambda^0(N,m,t)\,\Lambda^0(N,m,k)}\;.
\label{eq37}
\ee
Also, the $q^V$ from $V(k)$ is given by Eq. (\ref{eq19}) and the same equation gives $q^h$ from $H_0(t)$ by replacing $k$ by $t$. Now, some comments are in order. (i) Firstly let us comment on the role of $\lambda$ in Eq. (\ref{eq25}). It is well known that for very small values of $\lambda$, the strength functions will be delta functions at $\ek$ and as $\lambda$ increases they will take Breit-Wigner (BW) form. With further increase the BW form changes to $f_{CqN}$ with $\lambda$ sufficiently
large. Thus, for the strength functions to take $f_{CqN}$ form $\lambda$ needs to be large and this form will be good for $\xi^2\sim 1/2$ ($\xi^2=1/2$ is thermalization regime and then $\sigma^2_{H_0} = \lambda^2 \sigma^2_{V}$);
see \cite{EE7,EE8,zel10,qn-3,qn-12,qn-13}. An open question is to construct a $q$-BW form and also a form that
is applicable in the $q$-BW to $f_{CqN}$ transition region (this 
region also falls in the chaitoc domain with level fluctuations 
following GOE). (ii) It is seen from Eq. (\ref{eq33}) that $\gamma_1(\ek)$ is not zero and therefore $F_\kappa(E)$ from EE are not symmetrical. For $\ek$ negative $F_\kappa(E)$ will be skewed in the positive direction and for $\ek$ positive  it will be skewed in the negative direction . (iii) From Eq. (\ref{eq36}), it is easy to see that $\gamma_2(\ek)=q^{hv}(1-q^{hv})$ for $\ek=0$ and $\xi^2=1/2$ and hence in the thermodynamic region it is always positive. (iv) Numerical calculations show that for EE the $f_{CqN}$ form is certainly good for $|\ek| \le 2$ and it may need corrections beyond this (see Section \ref{sec43} for further discussion); (v) Using $f_{CqN}$ form it is possible to derive EE formulas for the quantum chaos measures information entropy and inverse participation ratio \cite{qn-3,qn-12,qn-13}. (vi) Though we do not have BEE formulas for the lower order moments $M_r(\ek)$, it is expected that the $f_{CqN}$ form applies to BEE in the dense limit (see \cite{qn-12} and Section \ref{sec43}).

\section{Further results from $q$-normal forms}
\label{sec4}

\subsection{Two-point correlation function with $q$-Hermite expansion}
\label{sec41}

Ensemble averaged (smoothed) eigenvalue density, transition strength density and strength functions that take $q$-normal forms as described in Section \ref{sec3} are directly useful in SSM (see Section \ref{sec5}). However, the errors for example in using the smooth eigenvalue density can be estimated only if we have the knowledge of two-point and higher order correlations in eigenvalues (topic of correlations in eigenvector components or transition strengths is a far more complex subject \cite{EE4}). More importantly,level and strength fluctuations as quantified by the two and
higher point functions are essential for understanding quantum chaos and thermalization in isolated finite many-particle quantum systems. Dyson defined correlation functions in eigenvalues \cite{Dyson} and the lowest of these is the two-point correlation function (smoothed eigenvalue density is the one-point function). For a random matrix, it is given by the ensemble average of the product of the density of eigenvalues at two eigenvalues say $E$ and $E^\pr$. Fluctuation measures such as the number variance and Dyson-Mehta $\Delta_3$ statistic  are defined by the two-point function and so also the variance of the level motion in the ensemble. As follows from the results in Section \ref{sec31}, clearly the eigenvalue density generated by a member of FEGOE($k$)/FEGUE($k$) can be expanded by starting with the (smoothed)
$q$-normal form and using the associated $q$-Hermite polynomials $He_\zeta(x|q)$. Covariances $\overline{S_\zeta S_{\zeta^\pr}}$ (overline representing ensemble average) of the expansion coefficients $S_\zeta$ with $\zeta \ge 1$ here determine the two-point function as they are a linear combination of the bivariate moments $\Sigma_{PQ}$ of the two-point function. Recently \cite{qn-15}, using binary correlation approximation formulas for
the covariances $\overline{S_\zeta S_{\zeta^\pr}}$ for low values of $\zeta$ are derived and they are briefly described in this Section. 
Let us stress that for the GOE/GUE not just the lowest covariances, but the full two-point function was obtained using the binary correlation approximation in \cite{EE4,FMP}. Using this method or any other, till now the two-point
function could not be obtained for FEGOE($k$)/FEGUE($k$) with $k \ne m$ (let us mention that $k=1$ is special \cite{Reta,Pand,Germ}). Here below,
we will restrict to FEGUE($k$) and the results can be extended to FEGOE($k$).

Given the ensemble averaged eigenvalue density $\overline{\rho(E)}$
of FEGUE($k$) where $\rho(E)$ is the eigenvalue density for each member of EGUE($k$), integral of $\rho(E)$ defines the distribution function,
$F(x) = d\,\int_{-\infty}^x \rho(E)\,dE$. Note that $F(x)$ gives number of levels up to the eigenvalue $x$ and $d=\binom{N}{m}$ is the matrix dimension. Now, the two-point correlation function $S^{\rho}(x,y)$ for the eigenvalues and its integral version $S^F(x,y)$ are (in this Section
$x$ and $y$ also denote energies or eigenvalues), 
\be
\barr{rcl}
S^{\rho}(x,y) & = & \overline{\rho(x)\;\rho(y)} -{\overline{\rho(x)}}\;{\overline{\rho(y)}}\;,\\
S^F(x,y) & = & d^2\; \dis\int_{-\infty}^x \dis\int_{-\infty}^{y} S^\rho(x^\pr ,y^\pr) dx^\pr dy^\pr \;\;=\;\;\overline{F(x)\;F(y)} -{\overline{F(x)}}\;{\overline{F(y)}}\;.
\earr \label{fl1}
\ee
It is clear that $S^\rho$ (and $S^F$) gives
measures for level fluctuations and the simplest two-point measure is the number variance $\Sigma^2(\on)$. Say, there are $n$ number of levels between energies $x$ and $y$. Then $n=F(x)-F(y)$ and $\on = \overline{F(x)} - \overline{F(y)}$. With these, the number variance 
$\Sigma^2(\on) = \overline{(n-\on)^2}$ is simply,
\be
\Sigma^2(\on) = S^F(x,x) + S^F(y,y) -2 S^F(x,y)\;.
\label{fl2}
\ee
In addition, the Dyson-Mehta $\Delta_3$ statistic is related to $\Sigma^2(\on)$ involving an integral with $\Sigma^2(r)$ \cite{EE4}.
Also, $S^F(x,x)$ gives the variance of the fluctuation in an eigenvalue $E$ measured in units of the local level spacing. Importantly, $S^F(x,y)$ and $S^\rho(x,y)$ can be probed or constructed using the bivariate moments ${\widetilde{\Sigma}}_{PQ}$ of $S^{\rho}(x,y)$,
\be
{\widetilde{\Sigma}}_{PQ} = \dis\int x^P\, y^Q \,S^{\rho}(x,y)
\,dx dy = \overline{\lan H^P\ran \lan H^Q\ran} -\overline{\lan H^P\ran}\;\;\overline{\lan H^Q\ran}\;.   
\label{fl3}
\ee
Also, with $\Sigma_{PQ} = \overline{\lan H^P\ran \lan H^Q\ran}$, we have $\Sigma_{P,0} = \overline{\lan H^p\ran^m}$, the $P$-th moment of $\overline{\rho(E)}$. 

Proceeding further, first the eigenvalue density $\rho(E)$ for various members of FEGUE($k$) can be expanded in terms of $q$-Hermite polynomials starting with $q$-normal giving, 
\be
\rho(E)\,dE = f_{qN}(\eh|q) \l[1+ \dis\sum_{\zeta \ge 1}^{\infty} S_\zeta\;\f{He_\zeta(\eh |q)}{\l[\zeta\r]_q!} \r]\,d\eh\;; \;\;\eh=(E-E_c)/\sigma\;.
\label{fl4}
\ee
Here, $S_\zeta$ are the expansion coefficients and the $S_\zeta$ should not be confused with $S^\rho(x,y)$ used for the two-point function. It is important to recall that the ensemble averaged eigenvalue density $\overline{\rho(E)}$ for FEGUE($k$) is $f_{qN}$, the $q$-normal. The $S_\zeta$'s in Eq. (\ref{fl4}) are for a given member of the FEGUE($k$) ensemble and it is easy to see that $\overline{S_\zeta}=0$. Note that Eq. (\ref{fl4}) gives an expansion for $S^\rho(x,y)$ in terms of $q$-Hermite polynomials (in the reminder of this paper, the symbols $x$ and $y$ are standardized variables), 
\be
S^{\rho}(x,y) = f_{qN}(x|q)\,f_{qN}(y|q)\,\dis\sum_{\zeta\,,\,\zeta^\pr = 1}^{\infty}
\overline{S_\zeta\,S_{\zeta^\pr}}\;\f{He_\zeta(x |q)}{\l[\zeta\r]_q!}\;\f{He_{\zeta^\pr}(y |q)}{\l[\zeta^\pr \r]_q!}\;.
\label{fl5}
\ee
Also we have easily,
\be
\lan H^p\ran = \overline{\lan H^p\ran} + \dis\sum_{\zeta \ge 1} S_\zeta\,\dis\f{\sigma^p}{\l[\zeta\r]_q!}\;\dis\int_{S(q)} \,x^p\,f_{qN}(x|q)\,He_\zeta(x|q)\,dx \;.
\label{fl6}
\ee
Note that $\sigma^2=\Sigma_{2,0}=\Sigma_{0,2}$. Now, writing $x^p$ in terms of $q$-Hermite polynomials and using the results in \cite{qn-8} (see also Section 2) we have the important result,
\be
\barr{l}
\lan H^p\ran = \overline{\lan H^p\ran} + \dis\sum_{\zeta \ge 1} S_\zeta\,\sigma^p\,C_{\f{p-\zeta}{2},p}(q)\;;\\
\Rightarrow 
{\hat{\Sigma}}_{PQ} = \dis\f{{\widetilde{\Sigma}}_{PQ}}{\l[\Sigma_{2,0}\r]^{(P+Q)/2}} = \dis\sum_{\zeta , \zeta^\pr = 1}^{\infty}
\overline{S_\zeta\,S_{\zeta^\pr}}\;C_{\f{P-\zeta}{2},P}(q)
C_{\f{Q-\zeta^\pr }{2},Q}(q)\;.
\earr \label{fl7}
\ee
The $\widetilde{\Sigma}_{PQ}$ is defined by Eq. (\ref{fl3}) and a formula
for the $C_{--}$ factors in Eq. (\ref{fl7}) is given in \cite{qn-8}. 
Let us add that ${\hat{\Sigma}}_{PQ}=0$ for $P+Q$ odd and similarly $\overline{S_\zeta\,S_{\zeta^\pr}}=0$ for $\zeta +\zeta^\pr$ is odd. Also, ${\hat{\Sigma}}_{P0}=0$, ${\hat{\Sigma}}_{PQ} =
{\hat{\Sigma}}_{QP}$, $\overline{S_\zeta}=0$ and  $\overline{S_\zeta\,S_{\zeta^\pr}}=\overline{S_{\zeta^\pr}\,S_{\zeta}}$. Using Eq. (\ref{fl7}) successively with $P+Q$ increasing from 2, the covariances $\overline{S_\zeta\,S_{\zeta^\pr}}$ can be written in terms of the moments ${\hat{\Sigma}}_{PQ}$. For example, formulas for $\zeta+\zeta^\pr \le 6$ are,
\be
\barr{rcl}
\overline{S_1\,S_1} & = & {\hat{\Sigma}}_{11}\;,\;\;\;
\overline{S_3\,S_1} =  {\hat{\Sigma}}_{31} - C_{13}\,{\hat{\Sigma}}_{11}\;,\;\;\;
\overline{S_2\,S_2} = {\hat{\Sigma}}_{22}\;,\\
\overline{S_5\,S_1} & = & {\hat{\Sigma}}_{51} - C_{15}\,\overline{S_3\,S_1} - C_{25}\,\overline{S_1\,S_1}\;,\;\;\;
\overline{S_4\,S_2} = {\hat{\Sigma}}_{42} - C_{14}\,\overline{S_2\,S_2}\;,\\
\overline{S_3\,S_3} & = & {\hat{\Sigma}}_{33} - C^2_{13}\,\overline{S_1\,S_1} - 2 C_{13}\,\overline{S_1\,S_3}\;.
\earr \label{fl8}
\ee
Here, $C_{1,3}=q+2$, $C_{1,4}=q^2+2q+3$, $C_{1,5}=q^3+2q^2+3q+4$ and $C_{2,5}(q) = q^3 + 3q^2 + 6q + 5$.

Formulas for the moments $\Sigma_{PQ}$ and hence for $\hat{\Sigma}_{PQ}$,
for a system of $m$ fermions in $N$ single particle states are derived, for $P+Q \le 8$ in \cite{qn-15} using the binary correlation approximation. For example,
\be
\barr{rcl}
\Sigma_{1,1} & = & \overline{\lan H\ran^m \lan H\ran^m} = \dis\f{1}{d^2}
\dis\sum_{\alpha_1, \alpha_2} \overline{H_{\alpha_1 \alpha_1}H_{\alpha_2 \alpha_2}} \\
\Sigma_{2,2} & = & \overline{\lan H^2\ran^m \lan H^2\ran^m} \\
&= & \l[\overline{\lan H^2\ran^m}\r]^2 + 2 \dis\f{1}{d^2} \dis\sum_{\alpha_1, \alpha_2,\alpha_a,\alpha_b}
\overline{H_{\alpha_1 \alpha_2} H_{\alpha_a \alpha_b}}\;\overline{H_{\alpha_2 \alpha_1} H_{\alpha_b \alpha_a}} \;.
\earr \label{fl9}
\ee
Note that here, $H_{\alpha \beta}$ are $H$ matrix elements in $m$ particle spaces. The expressions in Eq. (\ref{fl9}) are simplified by applying the Wigner-Racah algebra of $U(N)$ as described in \cite{un4,un5}. Then, we have the finite $N$ formulas, 
\be
\barr{rcl}
\hat{\Sigma}_{1,1} & = & \dis\f{\Lambda^0(N,m,m-k)}{\binom{N}{m}\,\Lambda^0(N,m,k)}\;, \\
\hat{\Sigma}_{2,2} & = & \dis\f{2\;\dis\sum_{\nu=0}^k\;\l[\Lambda^\nu(N,m,m-k)\r]^2 d(\nu)}{\l[\binom{N}{m}\,\Lambda^0(N,m,k) \r]^2}\;.
\earr \label{fl10}
\ee
Proceeding further, in \cite{qn-15} formulas are obtained for $\hat{\Sigma}_{P,Q}$ with $P+Q \le 8$. Then, in the asymptotic limit
defined by $N \rightarrow \infty$, $m \rightarrow \infty$, $m/N \rightarrow 0$ with $k$ finite, we have for example for $P+Q \le 6$
\be
\barr{rcl}
\sh_{1,1} & = & \dis\f{\binom{m}{k}}{\binom{N}{k}^2}\;,\;\;\;
\sh_{3,1} = 3 \dis\f{\binom{m}{k}}{\binom{N}{k}^2} = 3\,\sh_{1,1} \;,\;\;\;\sh_{2,2} = \dis\f{2}{\binom{N}{k}^2}\;,\\
\sh_{5,1} & = & 5 \dis\f{\binom{m}{k}}{\binom{N}{k}^2} \l[ 2 +
\,\dis\f{\binom{m-k}{k}}{\binom{m}{k}}\r] = (10+5q)\sh_{1,1} \;,\\
\sh_{4,2} & = & \dis\f{4}{\binom{N}{k}^2} \l[2 + \dis\f{\binom{m-k}{k}}{\binom{m}{k}}\r] = (4+2q)\sh_{2,2} \;,\\
\sh_{3,3} & = & 9 \sh_{1,1} + \dis\f{3}{\binom{m}{k}\,\binom{N}{k}^2} + O\l(\f{1}{\binom{N}{k}^4}\r) \;.
\earr \label{fl11}
\ee
Using these in Eq. (\ref{fl8}) show that for $q=1$, $\overline{S_\zeta\,S_{\zeta^\pr}} = \delta_{\zeta \zeta^\pr} \overline{S_\zeta^2}$. This
result was obtained in \cite{jbf10} as the assumption there is $q=1$ for FEGOE($k$) with $k << m$. However, we see from the formula for $q$ given before that for any reasonable values of $(N,m)$, the $q$ value will not
be close to $1$. In addition, we see that the formulas in Eqs. (\ref{fl8}) and (\ref{fl11}) (also, those for $P+Q=8$ given in \cite{qn-15}) agree with the formulas for GUE (i.e. for $k=m$ or $q=0$) given in \cite{EE4,FMP}. Also, they agree with the results for FEGUE($k$) in the $k^2/m \rightarrow 0$ limit as given in \cite{jbf10,EE4} and this corresponds to $q \rightarrow 1$ . Thus, the formulas in Eq. (\ref{fl11}) cover the two extreme limits and therefore expected to apply to all $k$ values. Let us add that direct derivation of asymptotic limit formulas for many other $\sh_{PQ}$ for $P+Q > 8$ may prove to be useful as they will provide systematics for $\hat{\Sigma}_{PQ}$ and hence for $\overline{S_iS_j}$. With this, it may be possible to carry out the sum in Eq. (\ref{fl5}) and obtain the two-point function (or the number variance) for FEGUE($k$) just as it was carried out using the moment method for GOE and GUE in the past \cite{EE4,EE7,FMP}.

\subsection{Ground state fluctuations with $q$ parameter}
\label{sec42}

Besides studying level fluctuations as described in the previous subsection, it is of interest to investigate how the distribution of the largest or smallest eigenvalue for FEGOE($k$) / BEGOE($k$) (also for the GUE versions) change as $q$ varies from $1$ to $0$ (i.e. as $k$ changes from 1 to $m$). This belongs to the subject of extreme value statistics (EVS) \cite{Majum-book}. It is well known that the classical EVS are classified into Frechet, Gumbel and Weibull distributions \cite{Gumbel-1958, Galambos-1987}. However, for the Wigner-Dyson GOE/GUE ensembles, EVS of the (lowest) largest eigenvalues is described by the celebrated Tracy-Widom (TW) distribution \cite{TW-1,TW-2,TW-3}. This corresponds to the situation with $k=m$ for FEGOE($k$)/ BEGOE($k$)/ FEGUE($k$)/ BEGUE($k$). Following these, there is a first attempt in \cite{EE8} to study the lowest eigenvalue distribution (LED) for {\it two-body} fermionic and bosonic EGOE, numerically  and they are found to follow the modified-Gumbel distribution that was used earlier in a Sherrington-Kirkpatrick model study in \cite{Pa-08}. As analytical study of EVS for FEGOE($k$)/BEGOE($k$) and FEGUE($k$)/BEGUE($k$) appears to be intractable, numerical analysis was attempted more recently for FEGOE($k$)/FEGUE($k$) in \cite{Be-23} and for BEGOE($k$)/BEGUE($k$) in \cite{qn-16}. For the fermionic ensembles, it is seen in the numerical calculations that the LED varies from Gaussian to TW form as $k$ changes from 1 to $m$. As attention was paid to the important result that the ensemble averaged eigenvalue density takes $q$-normal form for EE in the study of LED using bosonic ensembles in \cite{qn-16}, we will describe the results of the bosonic ensembles analysis briefly. Here, in the numerical studies Gaussian, TW and modified-Gumbel forms are employed.  

\begin{figure}
\centering
\includegraphics[width=0.6\linewidth]{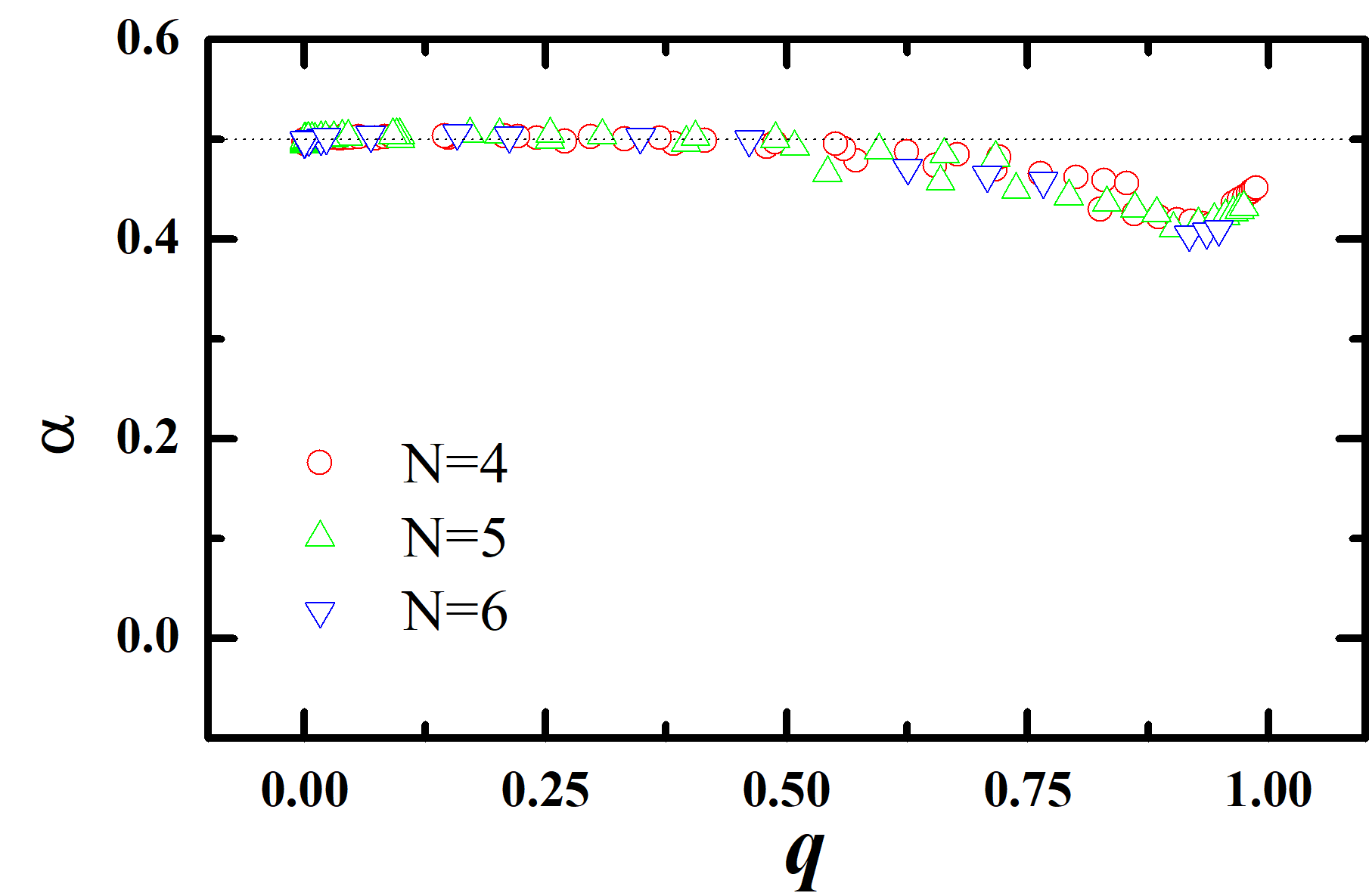}
\caption{Variation in the parameter $\alpha$ defined by the ansatz given in Eq. \eqref{eq-g1} for the centroids for the lowest eigenvalue distributions for a 1000 member BEGOE($k$) as a function of the parameter $q(N,m,k)$.  Used are: $m = 4-14$ for $N=4$, $m = 5-11$ for $N=5$ and $m =6-9$ for $N=6$.  In addition, $k = 1,  \;2, \ldots, \;m$. Though not shown in the figure, results for BEGUE($k$) are similar to those for BEGOE($k$). See text for more details. Figure is taken from \cite{qn-16}.}
\label{fig3-cent}
\end{figure}
\begin{figure}
\centering
\includegraphics[width=0.6\linewidth]{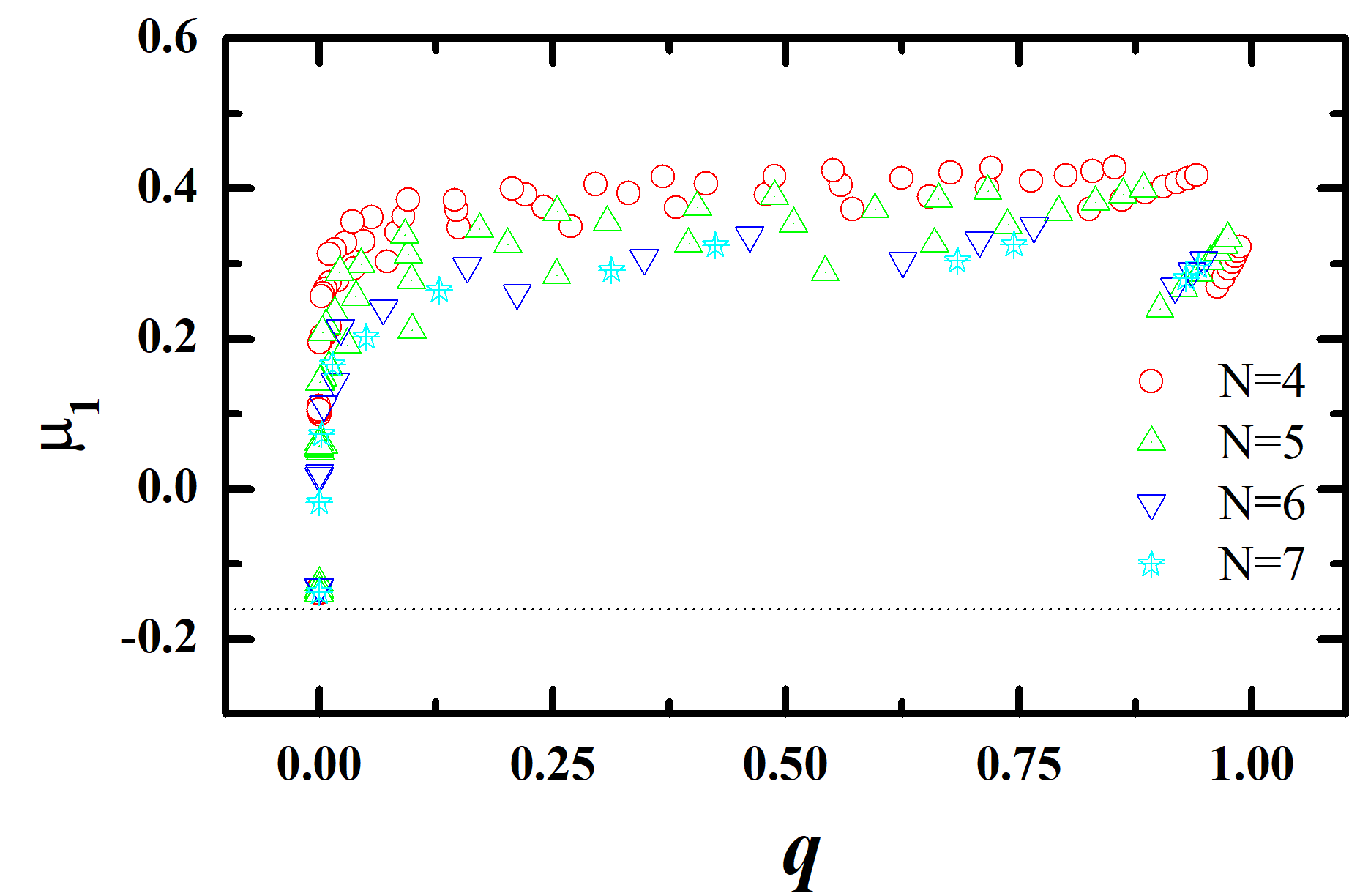}
\caption{Variation in the parameter $\mu_1$ defined by ansatz in Eq. \eqref{eq-g2} for the variances is shown for the LED for a 1000 member BEGOE($k$) as a function of the parameter $q(N,m,k)$. The $m$ values for $N=4$, 5 and 6 are as in Fig. \ref{fig3-cent} and for $N=7$ the $m$ values used are $7$ and $8$. In the calculations, $k$ is varied from $1$ to $m$.
Though not shown in the figure, results for BEGUE($k$) are similar to those for BEGOE($k$). See text and Ref. \cite{qn-16} for more details.}
\label{fig4-var}
\end{figure}

Firstly, the  $q$-normal distribution for the eigenvalue density shows that the distribution has a cut-off at 
$$
-2 \l[\beta \Lambda_B^{0}(N,m,k)\r]^{1/2} / \l[1-q(N,m,k)\r]^{1/2}
$$ 
at the lower edge when measured with respect to the ensemble averaged eigenvalue centroid; this follows easily from Eq.~\eqref{eq5}.  For finite $(N,m)$, there will be departures as $q$-normal is an asymptotic form for the eigenvalue densities. Therefore, the following parametrization was suggested in \cite{qn-16} for the centroid $\lambda_c$ of the lowest eigenvalues $\lambda$ for the various members of a FEGOE($k$)/FEGUE($k$). For a given $(N,m,k)$, the ansatz is,
\be
\lambda_c(N,m,k)  = \dis\frac{-2}{\dis\sqrt{1- q(N,m,k)}}{\l[ \beta \Lambda_B^{0}(N,m,k) \r]}^{\alpha}.
\label{eq-g1}
\ee
Note that for both BEGOE($k$) and BEGUE($k$), formulas for $q(N,m,k)$ and for $\Lambda^0_B(N,m,k)$ follow from Eq.~\eqref{eq20}. Also, as mentioned in Section \ref{sec3} the ensemble averaged variance of the eigenvalue density is $\Lambda^0_B(N,m,k)$. In the limit $q$-normal form is exact, the parameter $\alpha = 1/2$. Secondly, by applying Eq.~\eqref{eq20}, it is easy to see that Eq. \eqref{eq-g1} gives for $k=m$ the well-known result for TW for GOE/GUE. Therefore, using the calculated $\lambda$ values, via a least-square procedure, values of $\alpha$ for various $(N,m,k)$ values are obtained and the results are shown in Fig.~\ref{fig3-cent}. For $0 \leq q \leq 0.75$, the agreement with $q$-normal form result that $\alpha = 0.5$ is almost exact. Though not shown in the figure,  for $q(N,m,k) \gazz 0.8$ the deviations from $\alpha = 1/2$ are significant and they correspond to $k = 1$. As the member to member fluctuations in the spectral width are largest for $k=1$, corrections to the $k=1$ results are applied as described in \cite{qn-16} and the then the results obtained are shown in the figure. Thus, Eq. (\ref{eq-g1}) is a good formula for LED centroid.  

The variance for TW distribution is $D^{-1/6}$ for a $D$-dimensional GOE/GUE. This corresponds to BEGOE($k$)/BEGUE($k$) with $k = m$ and $D = \binom{N+k-1}{k}$. This is also proportional to the spectral width. Therefore, the following parametrization is suggested in \cite{qn-16} for BEGOE($k$)/BEGUE($k$), 
\be
\sigma_{\lambda} (N,m,k) = \left[\Lambda^{0}_B(N,m,k) \right]^{\mu_1},\;\;\;
\sigma_{\lambda}(N,m,k) = {\left[\Lambda^{0}_B(N,m,k) \right]}^{\mu_2} \binom{N+k-1}{k}^{-1/2}.
\label{eq-g2}
\ee
For $k=m$ case, i.e. for classical Gaussian ensembles or for TW, the exponents $\mu_1=-1/6$ and $\mu_2=1/3$ and they give correctly $\sigma_{\lambda}=D^{-1/6}$. Using the calculated $\lambda$ values for the same set of $(N,m,k)$ used in Fig. \ref{fig3-cent}, via least-square procedure,  the values of 
$\mu_1$ are obtained as a function of $q$ parameter and the results are shown in Fig.~\ref{fig4-var}. For small values of $q \lazz 0.1$,  there is a sharp increase in the values of $\mu_1$ and then, it essentially become a constant except for $k=1$ where the values of $\mu_1$ decrease.  Note that $\mu_1$ becomes positive which implies a larger variance compared to that for TW. This trend is common for both FEGOE/FEGUE \cite{Be-23}. Though not shown in the figure, the results for $\mu_2$ vs $q$ are essentially similar to those for $\mu_1$ vs $q$. Going beyond
the centroid and variance of LED, there are no ansatz formulas
for the shape parameters, skewness $S$ and kurtosis $\kappa$. Some numerical results for these are given in \cite{Be-23,qn-16}.
It is seen that in general, the skewness and kurtosis values are larger than those for TW distribution for intermediate $q$ values. For example, for BEGOE($k$) for $0.1 \lazz q \lazz 0.8$ the $S/S_{TW}$ decreases from $\sim 4$ to $2$ while $\kappa/\kappa_{TW}$ from $\sim 1.5$ to $1$ (note that $S_{TW} = 
-0.2935$ and $\kappa_{TW} = 3.1652$).  

\begin{figure}
\centering
\includegraphics[width=0.85\linewidth]{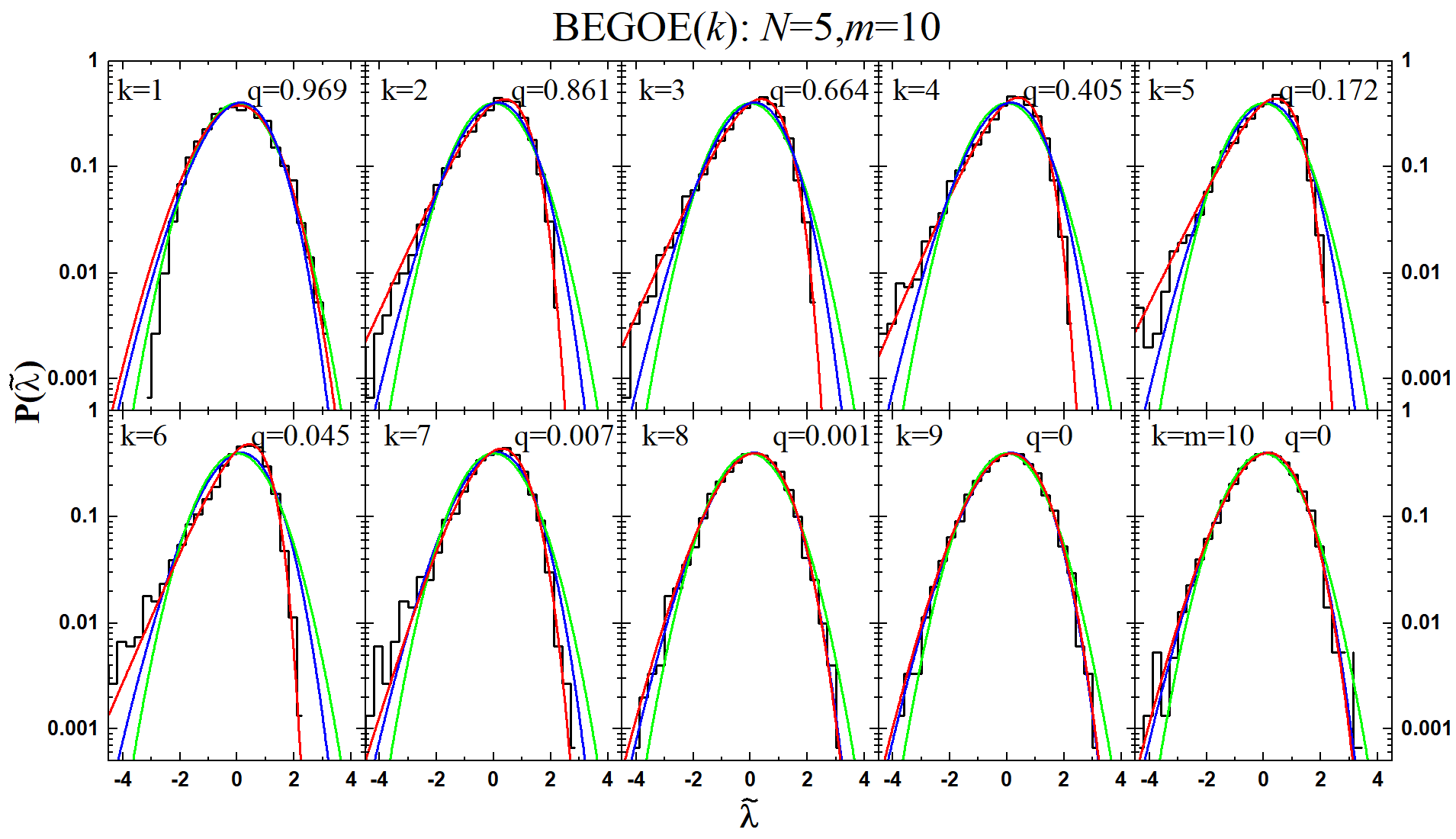}
\caption{Normalized probability distributions (histograms) $P({\tilde\lambda})$ for the lowest eigenvalues for a 5000 member BEGOE($k$) with $(N,m) = (5,10)$. The Gaussian (smooth green curves), modified Gumbel distribution $G_{\mu}(E)$ (smooth red curves) and the TW distribution (smooth blue curves) are superimposed in each panel with the numerical histograms. Figure is taken from \cite{qn-16}.}
\label{fig5-gs-be}
\end{figure}

Going beyond the lowest four moments, LED's obtained numerically are compared with Gaussian (${\cal G}$), classical TW and modified Gumbel distributions. Of these three, Gaussian is the simplest one and for the lowest eigenvalue $E$ with zero center and unit variance, we have ${\cal G}(E) = (2\pi)^{-1/2}\;\exp (-E^2/2)$. The TW form is given by an integral \cite{TW-1,TW-2,TW-3} and its numerical values are used to obtain a smooth form. The modified Gumbel distribution is given by \cite{Gumbel-1958, Galambos-1987},
\be
G_{\mu}(E)= w \exp \l[\mu\left(\frac{E-u}{v}\right)-\mu\exp \left(
\frac{E-u}{v}\right)\r]
\label{gumbel}
\ee
where $u$ and $v$ are rescaling parameters and $w$ is a normalization constant. The results for a 5000 member BEGOE($k$) are given in Fig. \ref{fig5-gs-be}. These are obtained for $N = 5$ and $m = 10$ system with $k$ varying from 1 to 10. The numerical histograms are computed using the scaled lowest eigenvalues $\tilde{\lambda} = [\sigma_\lambda(N,m,k)]^{-1}[\lambda - \lambda_c(N,m,k)]$. It is clearly seen from the figure that for $k=1$, the distributions are close to Gaussian form. However, for $2 \leq k \leq 6$ the distributions are close to the modified Gumbel form and there is clear transition to TW for $k = 10$.  Thus, the {\it LED for BEGOE($k$) changes from Gaussian to modified Gumbel to TW as $q(N,m,k)$ changes from $\sim 1$ ($k = 1$) to 0 ($k = m$)}. As the results in Figs. \ref{fig3-cent}, \ref{fig4-var} and \ref{fig5-gs-be} are obtained for modest values of $(N,m)$, these can be further substantiated (more importantly these may lead to an analytical formulation) by using larger $(N,m)$ values say with $N$ going to $10$ and $m$ upto $20$. Secondly, the results reported for fermionic systems \cite{Be-23} need to be re-analyzed using the $q$ parameter so that the results for fermionic and bosonic systems can be properly compared. Also, calculations for fermionic ensembles with much larger values of $(N,m)$ are needed. All these require
much larger computing power and these studies are for future. 

\subsection{Mean-field one-body part plus $k$-body interactions: Application of $q$-normal forms}
\label{sec43}

Nuclear Hamiltonians consist of a mean-field one-body part $h(1)$, a residual two-body part $V(2)$ and a small $3$-body part (perhaps also a four-body part) \cite{nn34}. Thus, $H=h+V$ with $h=h(1)$ and $V=V(2)$ or $V(2)+V(3)$ or $V(2)+V(3)+V(4)$. Then, FEGOE(1+2) or FEGOE(1+2+3) or FEGOE(1+2+3+4) needs to be analyzed. The FEGOE(1+2) has been analyzed in the past in detail,
assuming that the eigenvalue densities are close to Gaussian form, in terms of the additional parameter (relative strength of the two-body part) in the model \cite{EE7}. Following this, it is of interest to study the more general FEGOE(1+$k$)[also BEGOE(1+$k$)], i.e. with $H$ consisting of the mean-field one-body part and a $k$-body interaction, with $k$ changing from $2$ to $m$ and employing $q$-normal forms. This will be the same as considering $H$ in Eq. (\ref{eq25}) with $t=1$. Here below we will present some results for FEGOE(1+$k$) and BEGOE(1+$k$). We will return to FEGOE(1+2+3) and FEGOE(1+2+3+4) in Section \ref{sec5}.

Wavefunction structure for complex systems follow by examining  strength functions and quantum chaos measures such as number of principle components (NPC) and information entropy. Firstly, for interacting many-fermion systems (extension of all the results in this subsection to dense boson systems is straight forward) using Hamiltonian $H$, which is a sum of one-body $h(1)$ and an embedded GOE of $k$-body interactions $V(k)$ with strength $\lambda$, 
\be
H_{\lambda}(1+k) = h(1) + \lambda\,V(k),
\label{eqk1}
\ee 
it is easy to see that the eigenvalue density will be $q$-normal
for any $\lambda$ value (exceptions may happen for $h(1)$ generating singular spectra \cite{jbf3}). This result is well verified in a number of numerical examples in \cite{qn-12,qn-13,qn-4}. Going further, with $\lambda=0$, strength functions will be delta functions at the $h(1)$ basis state energies $E_\kappa$ with $\kappa$ denoting the $m$-particle $h(1)$ basis states. Without loss of generality, we consider $h(1)$ defined by a set of single particle energies $\epsilon_i$ with $i$ denoting sp states. Then,
$h(1)=\sum_i \epsilon_i n_i$ where $n_i$ is the number operator for the $i$-th sp state and say there are $N$ number of sp states. With $\lambda=0$, the $m$ fermion system basis states ($\kappa$)are given by the distribution of $m$ fermions in $N$ sp states. The basis state energies $E_\kappa=\sum_i \epsilon_i$ with the summation over the occupied orbits for the given $\kappa$ configuration (in practice, i.e. in numerical calculations, the diagonal energies $\lan \kappa \mid V(k) \mid \kappa\ran$ are added to $E_\kappa$). Now, increasing $\lambda$ value, the delta functions will spread and mix and thus, changing the form of the strength functions $F_{\kappa}(E)$ where $E$ are $H$ eigenvalues in $m$-particle spaces. For FEGOE(1+2), it is well established that the delta functions change to BW form after some value of $\lambda$ and with further sufficient increase in $\lambda$, the the BW form starts changing to near Gaussian form \cite{EE7,zel10}. Then, with further increase in $\lambda$ to a value $\lambda_t$ we reach the region of thermalization where different quantities defining the eigenstate properties such as entropy, strength functions, temperature etc, give the same values irrespective of the defining basis. The value of $\lambda_t$ is determined by the correlation coefficient $\xi$ in Eq. (\ref{eq31}). Then $\xi^2(\lambda_t) = \sigma^2_{h(1)}/\sigma^2_{h(1) + \lambda\,V(2)}=1/2$ \cite{AGK}. Assuming that this result extends to FEGOE(1+$k$), it follows from the results in Section 3.3 that the strength functions will take $f_{CqN}$ form when $\sigma^2_{h(1)}/\sigma^2_{h(1) + \lambda_t\,V(k)} \sim 1/2$ giving $\lambda_t=\sigma_{h(1)}/\sigma_{V(k)}$. Given 
the $\epsilon_i$, it is easy to write formulas for moments generated
by $h(1)$ \cite{jbf6,jbf3} and then it is easy to see that $\sigma^2_{h(1)} = [N(N-1)]^{-1} m(N-m) \sum_i {\tilde{\epsilon}}_i^2$. Note that $\tilde{\epsilon}_i = \epsilon_i - \overline{\epsilon}$; $\overline{\epsilon} = N^{-1}
\sum_i \epsilon_i$. With this and using $\sigma^2_{V(k)} = \Lambda^0(N,m,k)$ as given in Section \ref{sec31}, we have
\be
\lambda_t \sim \dis\sqrt{\dis\f{m(N-m) \dis\sum_i \tilde{\epsilon}_i^2}{N(N-1)\;\Lambda^0(N,m,k)}}\;.
\label{eqk2}
\ee
Thus, ensemble averaged strength functions for $H_{\lambda}(1+k)$ in many fermion spaces follow $f_{CqN}$ form for $\lambda \sim \lambda_t$, i.e. for sufficiently large $\lambda$ values. 

\begin{figure}
\centering
\includegraphics[width=0.8\linewidth]{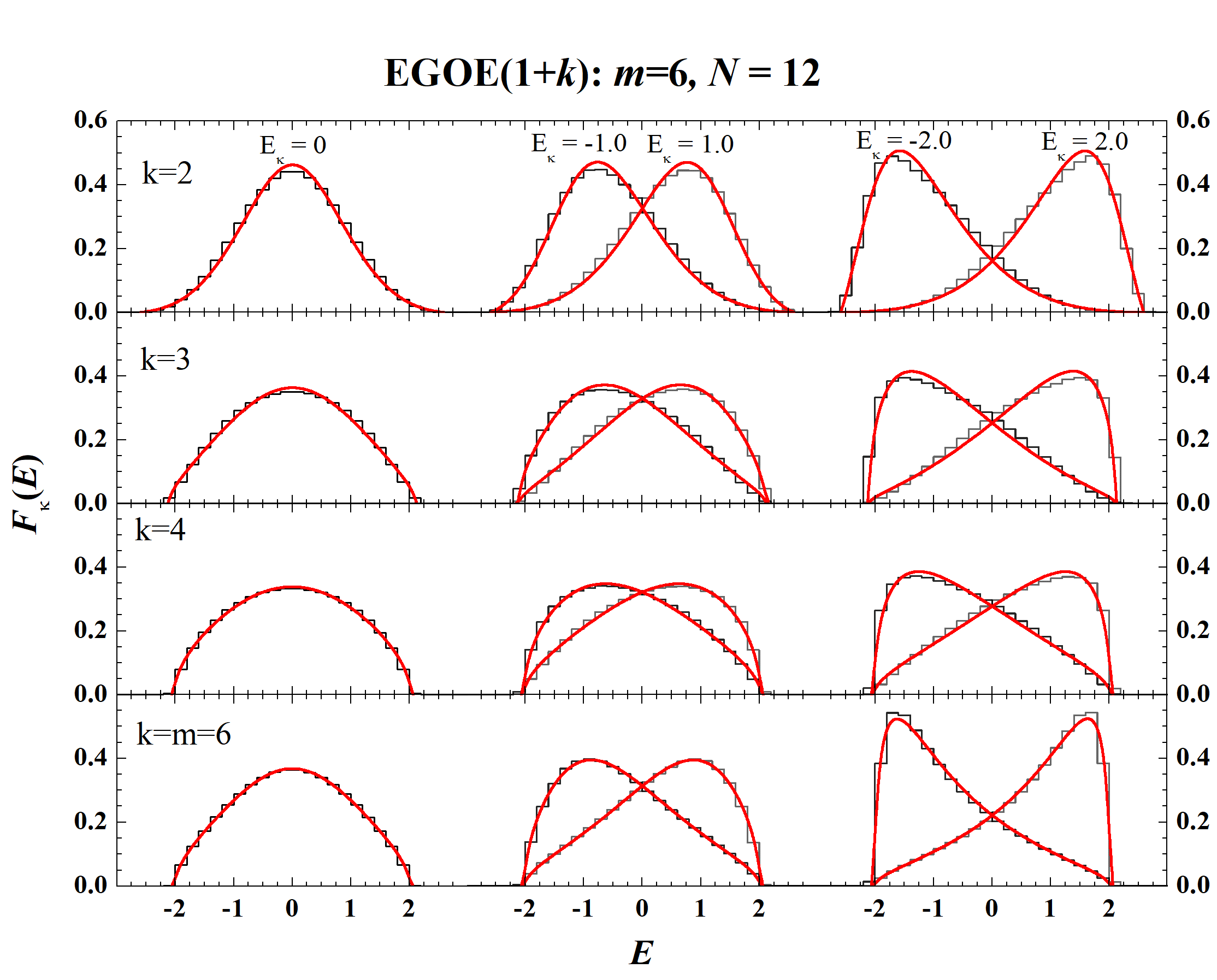}
\caption{Ensemble averaged strength function as a function of standardized energy $E$ (then $E=\eh$) for a 100 member FEGOE($1+k$) ensemble with $m = 6$ and $N = 12$. Results are shown for $k$=2,3,4 and $k = m=6$ and $k$-body interaction strength is chosen to be $\lambda=0.5$.  Histograms correspond to strength functions for $\ek=0$, $\pm 1.0$, and $\pm 2.0$. In the plots $\int F_{\kappa}(E) dE=1$. The smooth red curves are $f_{CqN}(\hat{E}|\ek;\xi,q)$ obtained using corresponding ensemble averaged values of $q$ and $\xi$. Figure is constructed using the results in \cite{qn-13}.}
\label{fig6-fke-f}
\end{figure}

Figure~\ref{fig6-fke-f} shows the ensemble averaged strength function results obtained for FEGOE($1+k$) with $m=6$ fermions in $N=12$ sp states. Here, $\lambda$ is chosen equal to 0.5, so that the system is in thermalization region for all $k$. 
In the ensemble calculations \cite{qn-13}, the $E_i$ and $E_{\kappa}$ spectra are scaled to have zero centroid and unit width for each member of the ensemble using the eigenvalue distribution and  $E_\kappa$-energies distribution, respectively. Therefore, $E \rightarrow \hat{E}$ and  $E_{\kappa}/\sigma_H=\xi\,\hat{E_{\kappa}}$.
The ensemble averaged $F_{\kappa}(E)$ results are shown for  $\hat{E_\kappa} = 0.0, \pm 1.0$, and $\pm 2.0$ using body rank $k = 2, 3, 4$ and $6$. All these
strength function histograms $F_\kappa(E)$ are fitted  with $f_{CqN}(\hat{E}|\ek;\xi,q)$.
For each $k$, the smooth curves in Figure~\ref{fig6-fke-f} are obtained using ensemble averaged values for
$\xi$ and $q$. It is clearly seen from the results shown in the figure that the ensemble averaged histograms are in very good agreement with the smooth forms obtained using $f_{CqN}$. From the results shown in the Figure~\ref{fig6-fke-f}, it is clearly seen that for sufficiently large $\lambda$, the smoothed strength functions $F_\kappa(E)$ are very well represented by $f_{CqN}$ and they make a transition from a near Gaussian form to the  semi-circle form, as the body rank $k$ in EGOE(1+$k$) changes from 2 to $m$. Also,  $F_\kappa(E)$ results for $\hat{E_\kappa}=0$ are symmetric and for $\hat{E_\kappa} \neq 0$, away from the center of the spectrum, $F_\kappa(E)$ results are asymmetrical about $\hat{E}$ as expected from Eq. (\ref{eq33}). 

\begin{figure}
\centering
\includegraphics[width=0.8\linewidth]{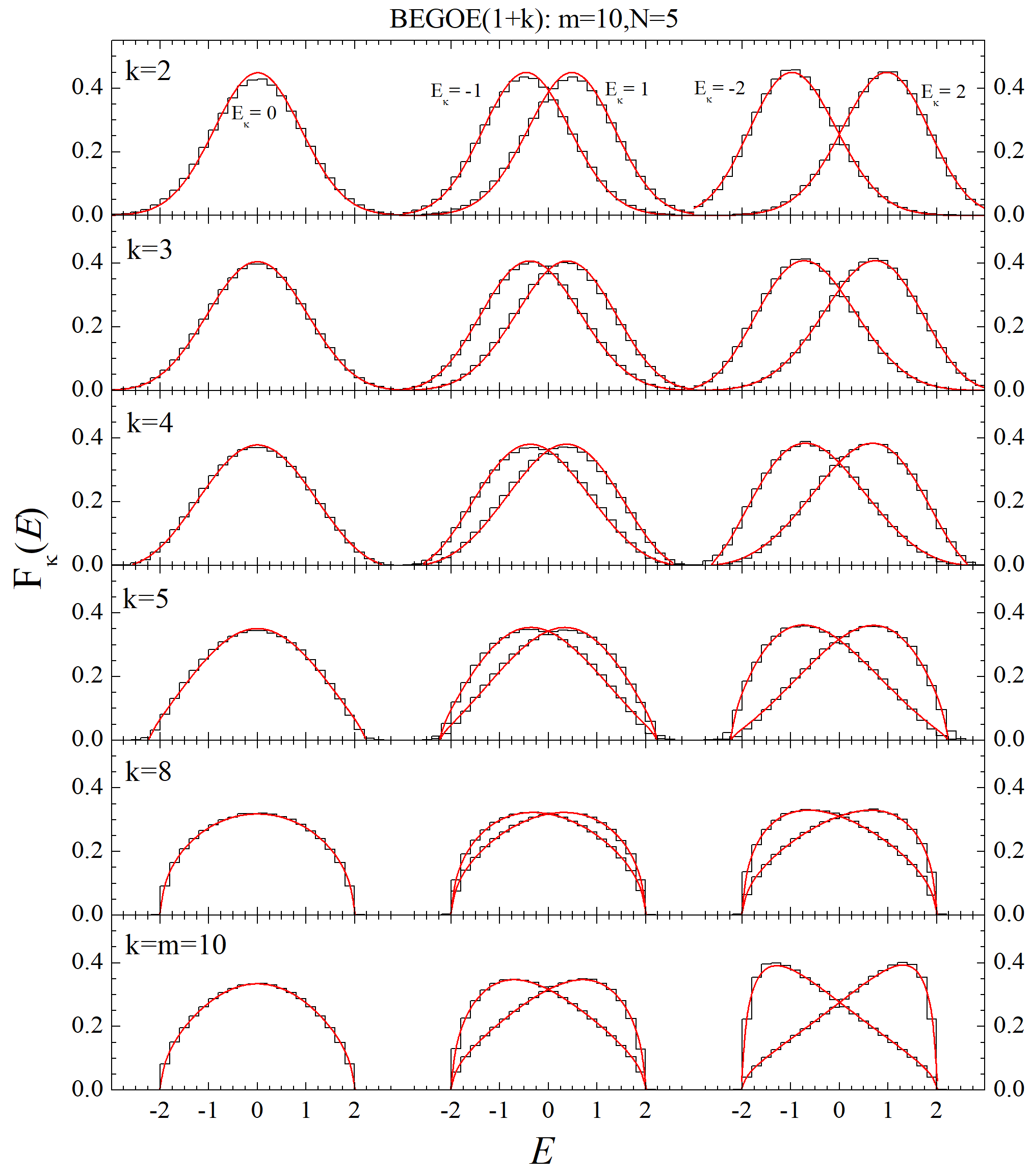}
\caption{Strength function vs. standardized energy $E$ (then $E=\eh$) for a system of $m = 10$ bosons in $N = 5$ sp states with $\lambda=0.5$ for different $k$ values in BEGOE(1+$k$) ensemble. An ensemble of 250 members is used for each $k$ and sp energies are chosen to be
$\epsilon_i=i+1/i$ (same as those used in Fig, 6). Histograms represent strength function plots obtained for $\ek=0$, $\pm 1.0$ and $\pm 2.0$. In the plots $\int F_{\kappa}(E) dE=1$. The continuous red curves are $f_{CqN}$, as in Fig. \ref{fig6-fke-f}, obtained using ensemble averaged values of $q$ and $\xi$. Figure is constructed from the results in \cite{qn-12}.}
\label{fig7-fke-b}
\end{figure}

All the results described above apply also to dense boson systems \cite{qn-12}. This is demonstrated in Figure~\ref{fig7-fke-b}. In this figure, histograms represent ensemble averaged $F_\kappa(E)$ results for a 250 member BEGOE(1+$k$) with $m=10$ bosons in $N=5$ sp states and $\lambda=0.5$. The strength function plots are obtained for $\ek = 0.0, \pm 1.0$ and $\pm 2.0$.  The value of $k$-body interaction strength is chosen such that $\lambda >> \lambda_t$, i.e. the system (for all $k$) exists in the region of thermalization \cite{Ch-PLA}. The histograms, representing BEGOE(1+$k$) results of strength functions, are compared with the conditional $q$-normal density function as given by, $F_\kappa(E)= f_{CqN}(x=E|y=E_\kappa;\xi,q)$. The smooth black curves in Figure~\ref{fig7-fke-b} for each $k$ are obtained via $f_{CqN}$ using corresponding ensemble averaged $\xi$ and $q$ values. The results in Figure~\ref{fig7-fke-b} clearly show very good agreement between the numerical histograms and continuous black curves for all body rank $k$. The $F_\kappa(E)$ results for $\ek=0$ are given in Figure~\ref{fig7-fke-b} clearly demonstrate that the strength functions are symmetric and also exhibit a transition from Gaussian form to semi-circle as $k$ changes from $m=2$ to $m=10$. The smooth form given by $f_{CqN}$  interpolates this transition very well. Going further, $F_\kappa(E)$ results for $\ek \neq 0$ are also shown in Figures~\ref{fig7-fke-b}. One can see that $F_\kappa(E)$ results are asymmetrical about $E$ as recognized  earlier for bosonic systems in \cite{CK2017}. Also, $F_\kappa (E)$ are skewed more in the positive direction for $\ek >0$ and skewed more in the negative direction for $\ek < 0$ exactly as predicted by Eq. (\ref{eq33}).

\begin{figure}
\centering
\includegraphics[width=0.8\linewidth]{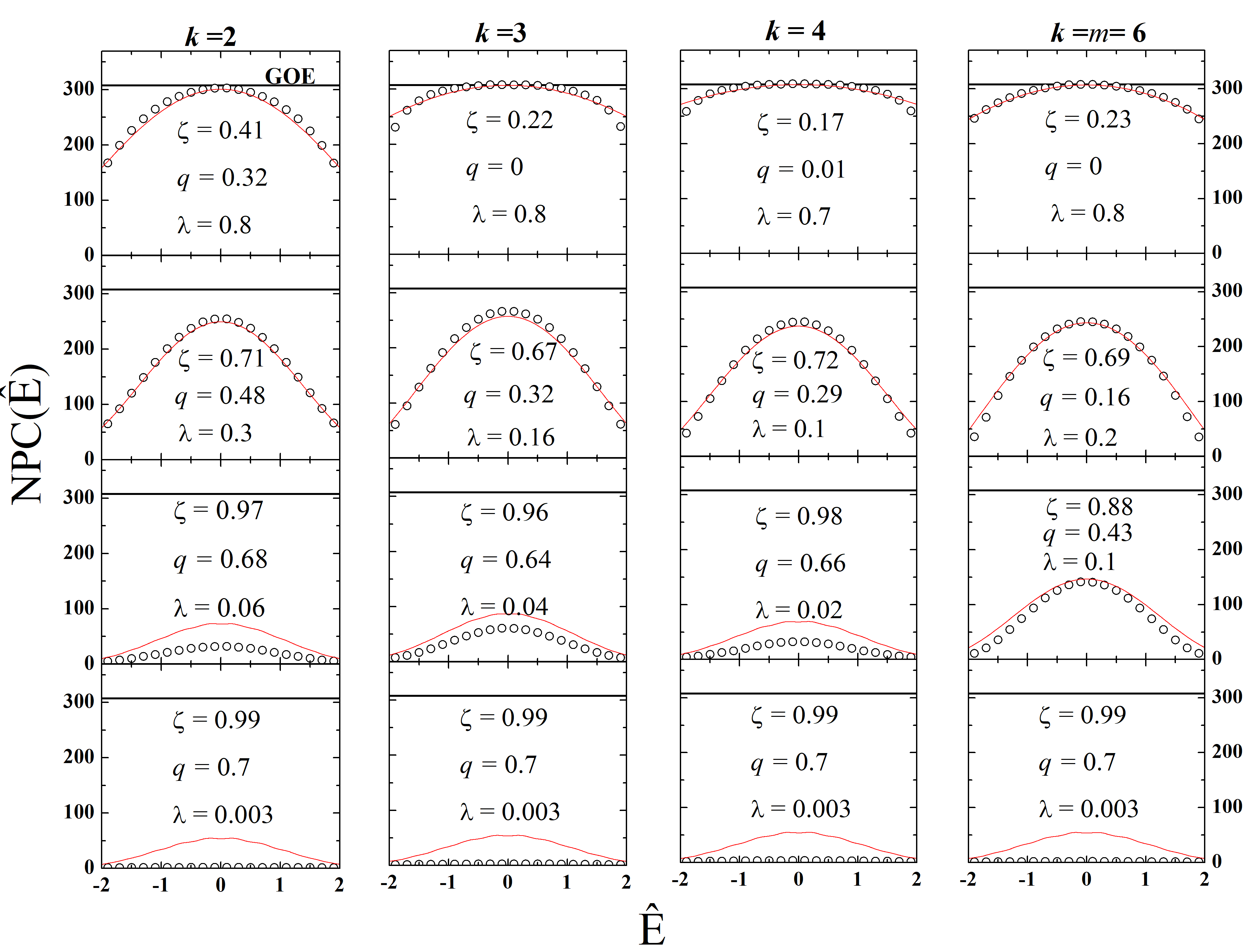}
\caption{Ensemble averaged NPC as a function of standardized energy $\eh$ for a 100 member FEGOE(1+$k$) ensemble just as in Fig. 6. Results are shown for $k=2$,$3$, $4$ and $6$ and for different values of the interaction strength $\lambda$. The circles are ensemble results and the smooth (red) curves are from Eq. (\ref{eqk5}). For each case, the values of $\xi$, $q$ and $\lambda$ are given in the figure. It is clearly seen that Eq. (\ref{eqk5}) is good when $\lambda$ value is large  and for small $\lambda$ values there are clear departures as here $q$-normal will not be good (strength functions will be close to BW or in BW to $f_{CqN}$ transition region). Figure is constructed from the results in \cite{qn-13}.}
\label{fig8-npc}
\end{figure}

Going beyond strength functions, we will briefly consider the quantum chaos measure NPC($E$) that gives number of basis states ($h(1)$ states) that make up an eigenstate with energy $E$. In terms of the $C$ coefficients in Eq. (\ref{eq26}), NPC is given by 
\be
NPC(E) = \l\{\dis\sum_\kappa \l|C^E_\kappa\r|^4\r\}^{-1}\;.
\label{eqk3}
\ee
Using the formulation developed for FEGOE(1+2) for deriving a formula for NPC in \cite{KoSa} and assuming that this extends to FEGOE(1+$k$), a formula in terms of an integral involving strength functions and state densities can be written. Then,
translating Eq.(4) in \cite{KoSa} to $q$-normal forms and assuming further that all the densities involved have the same $q$ value, we have
\be
NPC(E) = \dis\f{d}{3}\;\l\{\dis\int_{S(q)} d\ek\,f_{qN}(\ek) \l[f_{qN}(\eh)\r]^{-2}\;\l[f_{CqN}(\eh|\ek ; \xi,q)\r]^2\r\}^{-1}\;.
\label{eqk4}
\ee
Now, writing $f_{CqN}$ in terms of $q$-Hermite polynomials using Eqs. (\ref{eq13}), (\ref{eq9}) and (\ref{eq10}) in that order and then carrying out the $\ek$ integration using Eq. (\ref{eq7})
will give the formula,
\be
NPC(E) = \dis\f{d}{3}\;\l[h(\eh , \eh |\xi^2 , q)\r]^{-1}
\label{eqk5}
\ee
with the $h$ function given by Eq. (\ref{eq9}). Eq. (\ref{eqk5}) was given first in \cite{qn-12} with a summation over $q$-Hermite polynomials as in Eq. (\ref{eq10}). For a GOE $H$, we have $k=m$ with $\xi=0$. Then, Eq. (\ref{eqk5}) gives correctly NPC to be $d/3$ independent of $E$. With $f_{CqN}(\eh | \eh \; \xi^2 , q)$ for $q=1$ reducing to conditional normal form \cite{qn-6} and similarly $f_{qN}$ reducing to normal form, simplifying
$$
h(\eh, \eh|\xi^2 , q=1) = \l[f_{qN}(\eh)\r]^{-1}\;f_{CqN}(\eh | \eh ; \xi^2 , q=1)
$$
gives the formula
\be
NPC(E:q=1) = \dis\f{d}{3}\;\dis\sqrt{1-\xi^4}\;\exp-\f{\xi^2}{1+\xi^2} \,\eh^2\;.
\label{eqk6}
\ee 
This is same as the formula derived in \cite{KoSa} where $q=1$ is assumed. Thus, Eq. (\ref{eqk5}) gives correctly the known results for GOE and the formula in $q \rightarrow 1$ limit. In addition, it is well verified in a number of numerical ensemble calculations \cite{qn-12,qn-13} that for FEGOE(1+$k$) and BEGOE(1+$k$), Eq. (\ref{eqk5}) applies for all $k$ for sufficiently large values of the $\lambda$ parameter ($\lambda > \lambda_t$) appearing in Eq. (\ref{eqk1}). As an example we show in Fig. 8 some results for FEGOE(1+$k$).
 
\section{Statistical shell model with $q$-normal forms}
\label{sec5}

Nuclear Hamiltonians, as already mentioned in the previous Section, consist of a mean-field one-body part $h(1)$, a residual two-body part $V(2)$ and a small $3$-body part (perhaps also a four-body part). Just as SSM was developed and applied using Gaussian forms (see [1-10,15-22] in the past (assuming $H$ to be one plus two-body), with the $q$-normal forms now established to better approximate the state, transition strength and strength functions/partial densities, it is possible to develop SSM with $q$-normal forms. It is important to recognize that with $q$-normal forms, it is possible to consider $H$ to be not only one plus two-body but also one plus two plus three-body (may be plus four-body also). Following the first attempt in this direction in Ref. \cite{qn-4}, in this Section described are some basic approaches one may adopt using $q$-normal distributions and the associated $q$-Hermite polynomials in SSM. Note that they will include information about the fourth moment of the density of eigenvalues and the boundedness of the eigenvalue density and other distributions in a natural way in SSM. 

\subsection{Level densities}
\label{sec51}

Let us begin with nuclear level densities that are by definition statistical quantities and they are important as they are measurable and needed for Astrophysical reaction rates calculations. Given a nucleus with valence protons (say $m_p$ in number) occupying shell model sp orbits $j^p_1$, $j^p_2$, $\ldots$, and similarly $m_n$ number of valence neutrons occupying sp orbits $j^n_1$, $j^n_2$, $\ldots$, the $(m_p,m_n)$ space can be decomposed into $p-n$ configurations $(\tmp, \tmn)$ by distributing the nucleons in their respective valence sp orbits.
Then the state (eigenvalue) density $I^{(m_p, m_n)}(E)$ can be written as a sum of the partial densities defined over $(\tmp, \tmn)$ configurations giving, 
\be
I^{(m_p, m_n)}(E) = d(m_p,m_n)\,\rho^{(m_p, m_n)}(E) = 
\dis\sum_{(\tmp, \tmn)}\, d(\tmp,\tmn)\,\rho^{(\tmp, \tmn)}(E) = \dis\sum_{(\tmp, \tmn)}\,I^{(\tmp, \tmn)}(E) \;.
\label{eq.ann1}
\ee
Eq. (\ref{eq.ann1}) is exact and here the $d$'s are dimensions, $\rho$'s are normalized to unity and $I$'s are normalized to the dimensions. Without loss of generality, from now on, we will denote $(\tmp , \tmn)$ by $\wm$, $d(m_p,m_n)$ by $d$ and
$\rho^{m_p, m_n)}(E)$ by $\rho(E)$. The moments $M_p(\wm)$ of $\rho^{(\wm)}(E)$ are $M_p(\wm) = \lan H^p\ran^{\wm}$ with the centroid $\epsilon(\wm)=\lan H\ran^{\wm}$ and the variance $\sigma^2(\wm) =\lan H^2\ran^{\wm} -[\epsilon(\wm)]^2$. Now, using the $q$-normal form for $\rho^{(\wm)}(E)$ we have, assuming
$q$ is independent of $\wm$,
\be
I^{(m_p,m_n)}(E) = \dis\sum_{\wm} d(\wm) \rho^{(\wm)}(E) 
\approx  \dis\sum_{\wm}
\dis\f{d(\wm)}{\sigma(\wm)}\,f_{qN}^{(\wm)}(\eh(\wm)|q)
\label{eq.ann2}
\ee
with $\eh(\wm)=(E-\epsilon(\wm))/\sigma(\wm)$. The $f_{qN}$ is defined over the interval
$$
\l(\epsilon(\wm)-\dis\frac{2\,\sigma(\wm)}{\dis\sqrt{1-q}}\;,\;\epsilon(\wm)+\dis\frac{2\,\sigma(\wm)}{\dis\sqrt{1-q}}\r)\;.
$$
In practice, we can use the EGOE formula for $q$ for $p-n$ systems deducing via $\lan H^4\ran^{(m_p,m_n)}$ formula given in \cite{un5} and averaging its value for one, two and three-body
$H$. An alternative is to use $q$ as a free parameter. It is also possible to derive for a realistic $H$ the exact formula for $\lan H^4\ran^{(m_p,m_n)}$ using the methods given in \cite{jbf2,jbf6} but this is challenging with three-body forces.
As $\rho^{(\wm)}(E)$ is in fact a strength function, a better
approximation incorporating the results in Section \ref{sec33} is \cite{qn-4}
\be
\barr{l}
\rho^{(\wm)}(E) = \l\{\sigma(\wm)\r\}^{-1}\;f^{(\wm)}_{qN}(\eh |q) \l[1+\dis\frac{\gamma_1(\wm)}{[3]_q!} He_3(\eh|q) +\dis\f{(\gamma_2(\wm)+1-q)}{[4]_q!} He_4(\eh|q)\r]\;;\\
\\
\gamma_1(\wm) \approx -(1-q)\l[\dis\f{\epsilon(\wm)-E_c(m)}{\sigma(\wm)}\r]\;,\\
\\
\gamma_2(\wm) = (q-1) + (1-q)^2 \l[\dis\f{\epsilon(\wm)-E_c(m)}{\sigma_{\wm}}\r]^2 + (1-q^2) \dis\f{\sigma^2_h(m)}{\sigma^2_V(m)}\;.
\earr \label{eq.ann3}
\ee
It is possible to use Eqs. (\ref{eq.ann1})-(\ref{eq.ann3}) to calculate state densities. Alternatively, by replacing $\wm$ by
$\wm\,J$ everywhere will give level densities. However, calculating $J$ dependent centroids, variances and $q$ value, in particular with 3-body forces, is computationally intensive. Before going further, it is important to mention that Fig. 1 shows that $f_{qN}$ is good for $k$-body interactions and Fig. 6 shows that $f_{qN}$ is good also for $(1+k)$-body interactions. However, as we have $(1+2+3)$-body interactions in nuclei, we show in Fig. 9 an example demonstrating that $f_{qN}$ is good for these also.

\begin{figure}
\centering
\includegraphics[width=0.9\textwidth,height=0.85\textwidth]{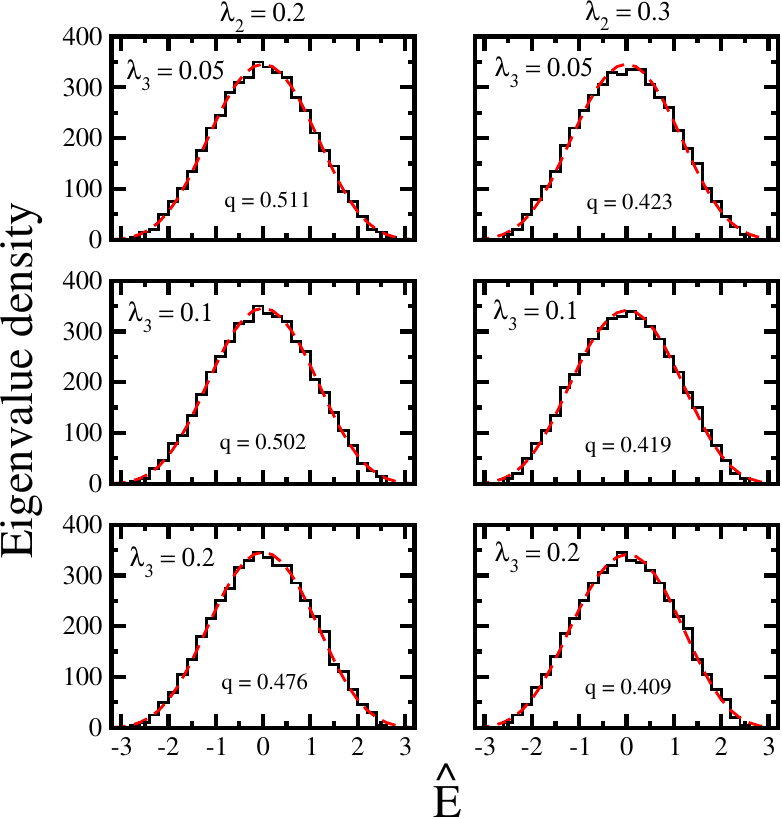}
\caption{Eigenvalue densities for a 1000 member FEGOE(1+2+3) ensemble with $H = h(1) + \lambda_2 V(2) +\lambda_3 V(3)$, where $\lambda_2$ and $\lambda_3$ are interaction strengths for two-body and three-body interactions respectively. Chosen is the system with $N = 12$ sp states and $m = 6$ fermions with $\lambda_2 = 0.2, \, 0.3$ and $\lambda_3 = 0.05, \, 0.1, \, 0.2$. The $V(2)$ and $V(3)$ are independent FEGOE's and $h(1)$ is defined by fixed sp energies $\epsilon_i = i + 1/i$; $i = 1, \, 2,\, \cdots,\, N$. Numerical results are shown as histograms and the analytical dashed curves are $f_{qN}(E)$ with $q$ values as given in the figure. Agreement between theory and numerical results is excellent. Figure is taken from \cite{qn-4}.}
\label{fig9-123}
\end{figure}
\begin{figure}
\centering
\includegraphics[width=0.875\textwidth,height=0.75\textwidth]{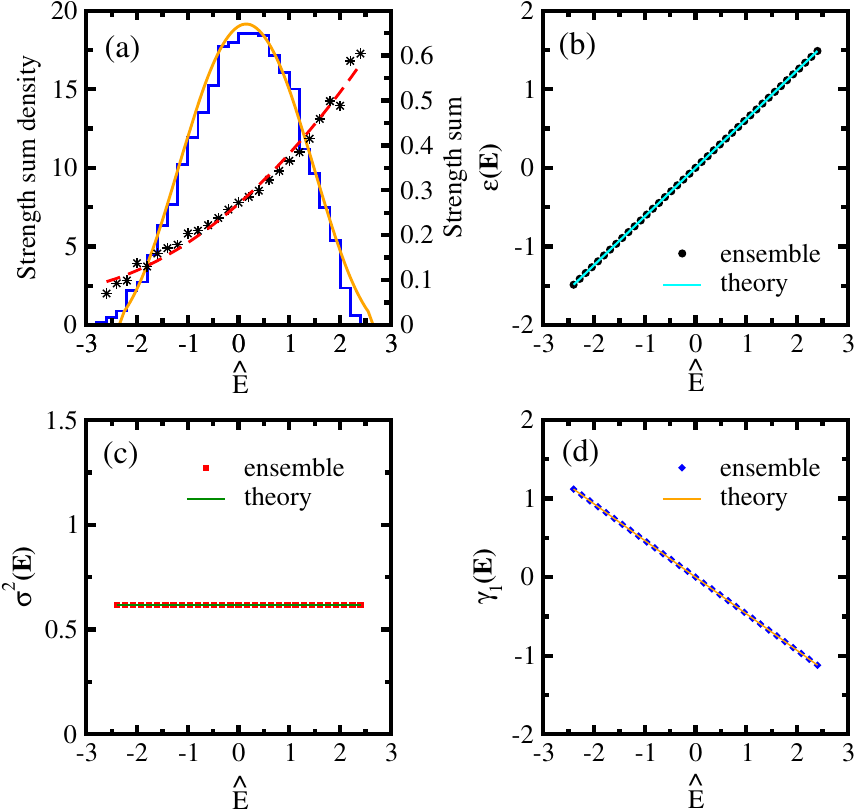}
\caption{(a) Transition strength sum and strength density, (b) centroid $\epsilon(E)$, (c) variance $\sigma^2(E)$ and (d) skewness $\gamma_1(E)$ for a one body transition operator $a^\dagger_2 a_9$ for an EGOE$(1+2+3)$ ensemble defined in Fig. 9. Numerical ensemble results (histogram and symbols such as stars, circles, squares and diamonds) are compared with analytical (smooth and dashed) curves.}
\label{fig10-strens}
\end{figure} 

\subsection{Transition strengths and strength sums}
\label{sec52}

Transition strengths and transition strength sums are in some situations measurable carrying new nuclear structure information and more importantly a theory for these is needed for many applications such as in calculating $\beta$-decay rates in Astrophysics, neutrinoless double $\beta$-decay transition matrix elements and so on. The SSM theory for transition strengths given in \cite{jbf9,jbf8,EE7} can be extended to incorporate the bivariate $q$-normal in place of bivariate Gaussian form. Also, the needed bivariate correlation coefficient for different types of transition operators follows from \cite{un5}. Some discussion of all these is given in \cite{qn-4}. 

Given a transition operator $\co$, a formula for transition strength sum $\lan \co^\dagger \co\ran^E$ (i.e. for the sum of the transition strengths originating from an eigenstate of $H=h+V$ with energy $E$) is, to a good approximation given by,
\be
\lan \cod \co\ran^{E} = \dis\sum_{(\tmp , \tmn)}\;\dis\f{I^{(\tmp ,\tmn)}(E)}{I^{(m_p ,m_n)}(E)}\;\lan \cod \co\ran^{(\tmp ,\tmn)}\;.
\label{eq.ann4}
\ee
Formulas for $\lan \co^\dagger \co\ran^{(\tmp ,\tmn)}$ can be written down for a variety of one and two-body operators \cite{jbf3}. With $\cod$ a creation operator, Eq. (\ref{eq.ann4}) gives shell model orbit occupancies and the formula is exact
for occupancies. Also, note that $\lan \cod \co\ran^{E} I^{(m_p,m_n)}(E)$ gives transition strength density.
In applying Eq. (\ref{eq.ann4}), the densities $I$'s will be
replaced by the corresponding $q$-normal densities as described
in Section \ref{sec51}.
Besides the non-energy strength sum,  also important are the lower order energy weighted strength moments, i.e. the centroid, variance and skewness, of the distribution of strengths originating from an eigenstate with energy E, as they are also measurable in many situations in nuclei. For example, the moments $M_p(E)$ are given by
\be
M_p(E) = \l\{\lan \cod \co\ran^E\r\}^{-1}\;\dis\sum_{E_f} |\lan E_f \mid \co \mid E\ran|^2\,(E_f)^p
\label{eq.ann5}
\ee
and then the centroid $\epsilon(E)=M_1(E)$ and the variance $\sigma^2(E) = M_2(E) -(M_1(E))^2$. Similarly, the skewness $\gamma_1(E)$ is defined via $M_3(E)$. As these are moments of the conditional density of the bivariate transition strength density, their variation with $E$ follows from Eqs. (31), (33) and (35). Then, $\epsilon(E)$ will be linear in $E$, the variance $\sigma^2(E)$ is a constant (does not depend on $E$) and $\gamma_1(E)$ will be linear in $E$ with a negative slope. All these results are tested in a numerical example and the results are shown in Figs. 10(a)-(d). In the calculations,  used is a EGOE$(1+2+3)$ ensemble with ($N = 12$, $m = 6$, $\lambda_2 = 0.3$ and $\lambda_3 = 0.2$). For the transition operator $\co$, chosen is the one-body operator $a^\dagger_2 a_9$. In Fig. 10(a), numerical results (histogram and stars) are compared with analytical curves for strength sum density and strength sum. They follow from Eq. (\ref{eq.ann5}) using $m$-particle averages. Then, the strength sum density is a marginal of the bivariate transition strength density and thus, it follows $q$-normal distribution and this corresponds to the smooth curve in the figure. Similarly, strength sum is the ratio of strength sum density and state density. As the transition operator is not completely random and the Hamiltonian operator has a fixed one-body part along with a mixture of two and three body rank operators, there is a shift of the centroid of the strength sum density relative to the state density centroid. Going to Figs. 10(b)-(d), it is clearly seen that the strength centroid $\epsilon(E)$, variance $\sigma^2(E)$ and skewness $\gamma_1(E)$ follow from the equations for the moments of the conditional $q$-normal distribution; see  Eqs. (31), (33) and (35). The agreements with theory are very good with some deviations at the spectrum edges.

In conclusion, it is important to add that the Gaussian form used in the past in SSM is reasonably good as long as the systems considered have a sufficiently large number of particles and the Hamiltonian is $(1+2)$-body. However, with growing knowledge on 3-body (perhaps also 4-body) interactions in nuclei, certainly in the future one needs the formulation, with $q$-normal forms, as briefly described in this Section. In future, it is important to carry out tests of SSM with $q$-normal forms using shell model codes with realistic $(1+2+3)$-body Hamiltonians. In addition, it is also important to carry out applications to nuclear level densities, astrophysical reaction rates calculations and so on. 
 
\section{Conclusions}
\label{sec6}

Embedded random matrix ensembles, introduced 50 years ago, continue to be of interest in nuclear physics in particular and in quantum many-particle physics in general. The EE provide the basis for SSM approach for nuclear structure. More recently (from 2017), following the RMT results, for the so called SYK model involving Majorana fermions, as derived by Verbaarschot and collaborators, a new direction in exploring EE and SSM has opened. Using the formulas for the lower order moments of eigenvalue densities, transition strength densities and strength functions on one hand and a variety of numerical calculations on the other, it is now well established that FEE($k$) and BEE($k$) indeed generate $q$-normal forms. These results are described in Sections \ref{sec31}-\ref{sec33}. The $q$-normal forms and some of their properties are given in Section \ref{sec2} for easy reference. Further,
presented are some results, in Sections \ref{sec41} and \ref{sec42} showing the role of the $q$ parameter in level fluctuations in the bulk and also in ground state (lowest eigenvalue) fluctuations. Although
analytical results showing $q$-normal forms are available only for FEE($k$) and BEE($k$), in nuclear physics applications in particular (for SSM), it is important to consider $k$-body interactions in the presence of a mean-field one-body term. As shown
using numerical examples in Section \ref{sec43}, with strong enough $k$-body interaction strength, the EE(1+k) also generate $q$-normal forms for various densities. Further, the $q$-normal forms are also good with the more realistic $(1+2+3)$-body interactions (see Figs. 9 and 10). 

With the $q$-normal forms well established, clearly it is necessary to modify SSM formulations used before by replacing Gaussians with $q$-normal forms. Some aspects of SSM with $q$ normal forms and the associated $q$-Hermite polynomials are
presented in Section \ref{sec5} with more details given in \cite{qn-4}.
In applying SSM with $q$ normal forms, a technical problem that needs to be solved is in extending the codes in \cite{zel6} to calculate also $\lan H^4\ran^{(\tmp , \tmn)J}$ or at least $\lan H^4\ran^{(m_p m_n)J}$. These will give the $q$ parameter values
with realistic nuclear interactions. Another important issue that needs to be addressed (this will extend the scope of SSM) is to solve the embedded random matrix that includes  multi-$\hbar\omega$ mixing. For this one has to consider for example the partitioned FEGOE described in Section 13.3 and Fig. 13.3 in \cite{EE7}. It is expected that these ensembles will give multi-modal distributions. Here, it is important to mention that after the NTSE-2026 meeting, there appeared a preprint presenting a more formal derivation of the $q$-normal forms for FEE($k$) and BEE($k$) using a method based on 'Wick product of non-commuting Gaussian random variables' \cite{JTST}. It remains to be seen if this new method gives a solution to the partitioned FEGOE and also solve the two-point correlation function describing level fluctuations.
We hope that this review will lead to further investigations of EE and SSM and their applications. Clearly, these future explorations and applications need HPC (High-Performance Computing).

\section*{Acknowledgments}

Thanks are due to S. Tomsovic for some useful correspondence. 
NDC acknowledges financial support from University Research
Project~No.DR/Dir./26-27/17/Sr~No-5.

\ed